\pdfoutput=1
\documentclass[sigconf, nonacm]{acmart}

\makeatletter
\def\@ACM@checkaffil{
	\if@ACM@instpresent\else
	\ClassWarningNoLine{\@classname}{No institution present for an affiliation}%
	\fi
	\if@ACM@citypresent\else
	\ClassWarningNoLine{\@classname}{No city present for an affiliation}%
	\fi
	\if@ACM@countrypresent\else
	\ClassWarningNoLine{\@classname}{No country present for an affiliation}%
	\fi
}
\makeatother

\usepackage{enumitem}
\usepackage{algorithm, algorithmic}
\usepackage{multirow}
\usepackage{subfigure}
\usepackage{bbding}
\usepackage{makecell}
\usepackage{pifont}
\usepackage{flushend}

\newcommand\vldbdoi{XX.XX/XXX.XX}
\newcommand\vldbpages{XXX-XXX}
\newcommand\vldbvolume{14}
\newcommand\vldbissue{1}
\newcommand\vldbyear{2020}
\newcommand\vldbauthors{\authors}
\newcommand\vldbtitle{\shorttitle} 
\newcommand\vldbavailabilityurl{https://anonymous.4open.science/r/A-ChatGPT-based-paradigm-for-detecting-bugs-in-graph-database-engines-3C56/}
\newcommand\vldbpagestyle{plain}

\begin{document}
	
	\title{Effective Bug Detection in Graph Database Engines: An LLM-based Approach}
	
	\author{Jiayi Wu}
	\affiliation{%
		\institution{Beijing Institute of Technology}
		\city{Beijing}
		\state{China}
	}
	\email{wjybit97@outlook.com}
	
	\author{Zhengyu Wu}
	\affiliation{%
		\institution{Beijing Institute of Technology}
		\city{Beijing}
		\state{China}
	}
	\email{jeremywzy96@outlook.com}
	
	\author{Rong-Hua Li}
	\affiliation{%
		\institution{Beijing Institute of Technology}
		\city{Beijing}
		\state{China}
	}
	\email{lironghuabit@126.com}

	\author{Hongchao Qin}
	\affiliation{%
		\institution{Beijing Institute of Technology}
		\city{Beijing}
		\state{China}
	}
	\email{qhc.neu@gmail.com}

	\author{Guoren Wang}
	\affiliation{%
		\institution{Beijing Institute of Technology}
		\city{Beijing}
		\state{China}
	}
	\email{wanggrbit@gmail.com}

	\begin{abstract}
		Graph database engines play a pivotal role in efficiently storing and managing graph data across various domains, including bioinformatics, knowledge graphs, and recommender systems. Ensuring data accuracy within graph database engines is paramount, as inaccuracies can yield unreliable analytical outcomes. Current bug-detection approaches are confined to specific graph query languages, limiting their applicabilities when handling graph database engines that use various graph query languages across various domains. Moreover, they require extensive prior knowledge to generate queries for detecting bugs. To address these challenges, we introduces DGDB, a novel paradigm harnessing large language models(LLM), such as ChatGPT, for comprehensive bug detection in graph database engines. DGDB leverages ChatGPT to generate high-quality queries for different graph query languages. It subsequently employs differential testing to identify bugs in graph database engines. We applied this paradigm to graph database engines using the Gremlin query language and those using the Cypher query language, generating approximately 4,000 queries each. In the latest versions of Neo4j, Agensgraph, and JanusGraph databases, we detected 2, 5, and 3 wrong-result bugs, respectively.
		
	\end{abstract}
	
	\maketitle
	\pagestyle{\vldbpagestyle}
	\begingroup\small\noindent\raggedright\textbf{PVLDB Reference Format:}\\
	\vldbauthors. \vldbtitle. PVLDB, \vldbvolume(\vldbissue): \vldbpages, \vldbyear.\\
	\href{https://doi.org/\vldbdoi}{doi:\vldbdoi}
	\endgroup
	\begingroup
	\renewcommand\thefootnote{}\footnote{\noindent
		This work is licensed under the Creative Commons BY-NC-ND 4.0 International License. Visit \url{https://creativecommons.org/licenses/by-nc-nd/4.0/} to view a copy of this license. For any use beyond those covered by this license, obtain permission by emailing \href{mailto:info@vldb.org}{info@vldb.org}. Copyright is held by the owner/author(s). Publication rights licensed to the VLDB Endowment. \\
		\raggedright Proceedings of the VLDB Endowment, Vol. \vldbvolume, No. \vldbissue\ %
		ISSN 2150-8097. \\
		\href{https://doi.org/\vldbdoi}{doi:\vldbdoi} \\
	}\addtocounter{footnote}{-1}\endgroup
	
	\ifdefempty{\vldbavailabilityurl}{}{
		\vspace{.3cm}
		\begingroup\small\noindent\raggedright\textbf{PVLDB Artifact Availability:}\\
		The source code, data, and/or other artifacts have been made available at \url{\vldbavailabilityurl}.
		\endgroup
	}

	\section{Introduction}
	In the contemporary information age, a vast volume of data is generated and accrued, frequently characterized by intricate relationships and connections, and such data are often represented as graph structures. While relational databases are designed based on a tabular model\cite{1} with Schema Rigidity, making it difficult to handle complex graph data. In response to this constraint, graph database engines have been designed to facilitate the storage and retrieval of graph data. They use the structure of nodes and edges to efficiently represent and process semi-structured data such as various relationships and networks\cite{2,3,4}. Graph database engines like Neo4j\cite{6}, JanusGraph\cite{7}, AgensGraph\cite{11}, TinkerGraph\cite{10}, OrientDB\cite{8}, and HugeGraph\cite{9}, which rank prominently in the DB-Engine Ranking list\cite{5}, are widely utilized across various domains. For instance, on social media platforms such as Facebook and Twitter, graph database engines are employed to manage and analyze social relationships and behavioral patterns among users, aiding in the detection of fake users and online fraudulent activities\cite{14}. In recommender systems, the utilization of graph database engines can dynamically generate real-time recommended content based on users' most recent actions and social connections\cite{12,13}. In the field of logistics and freight transportation, graph database engines can be harnessed to determine optimal delivery routes,  thereby reducing costs and enhancing efficiency\cite{15,16}.
	
	\begin{table*}[t]
		\caption{Comparison of the proposed DGDB with closely related works.}
		\label{tab:commands}
		\footnotesize
		\begin{tabular}{ccccc}
			\toprule
			Method& Category&Language&\makecell{Not require large prior knowledge}&\makecell{High proportion of  \\non-empty-result queries}\\
			\midrule
			Grand\cite{18} &\makecell{the method applying
				the same query\\ on various graph database engines} &gremlin&\small\color{red}{\XSolidBrush}& \small\color{red}{\XSolidBrush}\\ \\
			GDSmith\cite{19} &\makecell{the method applying
				the same query\\ on various graph database engines}& cypher & \small\color{red}{\XSolidBrush} &\small\color{green}{\Checkmark}\\ \\
			PP-DB\cite{17} & \makecell{the method for utilizng various queries to detect bugs\\ within the same graph database engines}& all & \small\color{red}{\XSolidBrush} &\small\color{red}{\XSolidBrush}\\ 
			\midrule
			DGDB&\makecell{the method applying
				the same query\\ on various graph database engines}& all&\small\color{green}{\Checkmark}&\small\color{green}{\Checkmark}\\
			\bottomrule
		\end{tabular}
	\end{table*}

	As shown in Table 1, existing methods for detecting bugs in graph database engines can be broadly divided into two categories: methods for utilizng various queries to detect bugs within the same graph database engines, methods applying the same query on various graph database engines. The former typically involves detecting bugs in graph database engines by generating equivalent queries\cite{36} or utilizing predicate partitioning\cite{17,35,37}. The latter approach entails running the same queries on different graph database engines and identifying bugs based on discrepancies in query results\cite{18,19}. The former method demands users with substantial prior knowledge on the graph model and query objectives when generating equivalent queries to identify bugs in the graph database engine\cite{36}. In the predicate partitioning approach for detecting graph database engine bugs\cite{17}, it starts by constructing queries based on a top-down expression generator\cite{20}, and then apply Ternary Logic Partitioning techniques to create correspondent queries, namely the "True, False or Null" queries, for running on graph database engine. Such a method assess whether the subset of the final query results intersect to detect graph database engine bugs. While this approach successfully identifies some logical bugs, its effectiveness is limited due to the lower quality of queries generation and the manual need for reducing test samples.	
	
	Most current research focuses on the latter approach and the most widely adopted method is running randomly generated queries \cite{20,21,22} on different graph database engines for bug detection. However, the generated queries often return empty results in many samples, which compromise the efficiency for bug detection. To address this issue, Grand\cite{18} proposed a method based on traversal models to generate gremlin queries\cite{34}, significantly increasing the non-empty-result query ratio. However, this method is only applicable to gremlin-based graph database engines and requires users to master the gremlin API\cite{43}. GDsmith\cite{19} proposed a method to generate cypher queries\cite{33} based on the guidance of property graphs. This method first generates a cypher skeleton, then creates patterns based on graph information, and ensures, through static query analysis, that the generated queries have high non-empty-result query ratios, thereby improving the efficiency of bug detection. However, this method is only applicable to cypher-based graph database engines and requires proficient mastery of cypher language.

	Large language models(LLM)\cite{39}, as one of the most popular research directions in the field of artificial intelligence, are trained on massive text data, enabling models to understand and generate diverse text content, fostering innovation in many domains. For instance, OpenAI's ChatGPT model\cite{38}, which is currently the most widely used large language model, is used in the recommendation domain to enhance text content, combining text with original content to improve the effectiveness of recommendation\cite{25}; in the financial domain, it is used to infer network structures from text data and then integrate them with graph neural networks\cite{40} to effectively predict stock trends\cite{26}. Meta proposed the SAM model\cite{23, 41}, which can output specified image regions based on prompt information and has been applied in the fields of medical image segmentation and video object segmentation.  Peking University's AIGC Lab introduced the ChatLaw model\cite{24}, which not only provides legal consulting services to the general public but also serves as an assistant to professional lawyers. 
	
	Based on ChatGPT's natural language understanding and natural language generation capabilities, we propose a simple paradigm for detecting bugs in graph database engines. Compared to existing bug detection methods, this paradigm does not require extensive prior knowledge but can automatically detect bugs in graph database engines using different graph query languages, demonstrating extremely high applicability. Additionally, benefiting from the diverse and complex queries generated by this paradigm, it can detect wrong-result bugs in the latest versions of graph database engines that prove challenging for existing bug detection methods to identify. Specifically, we start by randomly generating a property graph, then input the graph data, query generation instructions, and query generation constraints into ChatGPT to generate queries that meet these conditions. Finally, we run the queries on different graph database engines and detect bugs based on the discrepancies in query results.

	This paper has made the following contributions:
	\begin{itemize}[leftmargin=*]
		
		\item We designed a prompt template that helps ChatGPT generate a high proportion of non-empty result queries and enhances the diversity of generated queries, improving the efficiency of detecting bugs.
		
		\item We proposed a simple paradigm for detecting bugs based on ChatGPT, which does not require a significant amount of prior knowledge and can be applied to graph database engines using different graph query languages.
		
		\item We applied this paradigm to graph database engines using Cypher and Gremlin languages, effectively detecting 7 and 3 wrong-result bugs, respectively, demonstrating the effectiveness and applicability of the proposed paradigm.
		
	\end{itemize}
	
	\begin{figure}[t]
		\centering
		\includegraphics[width=6.9cm, height=5.4cm]{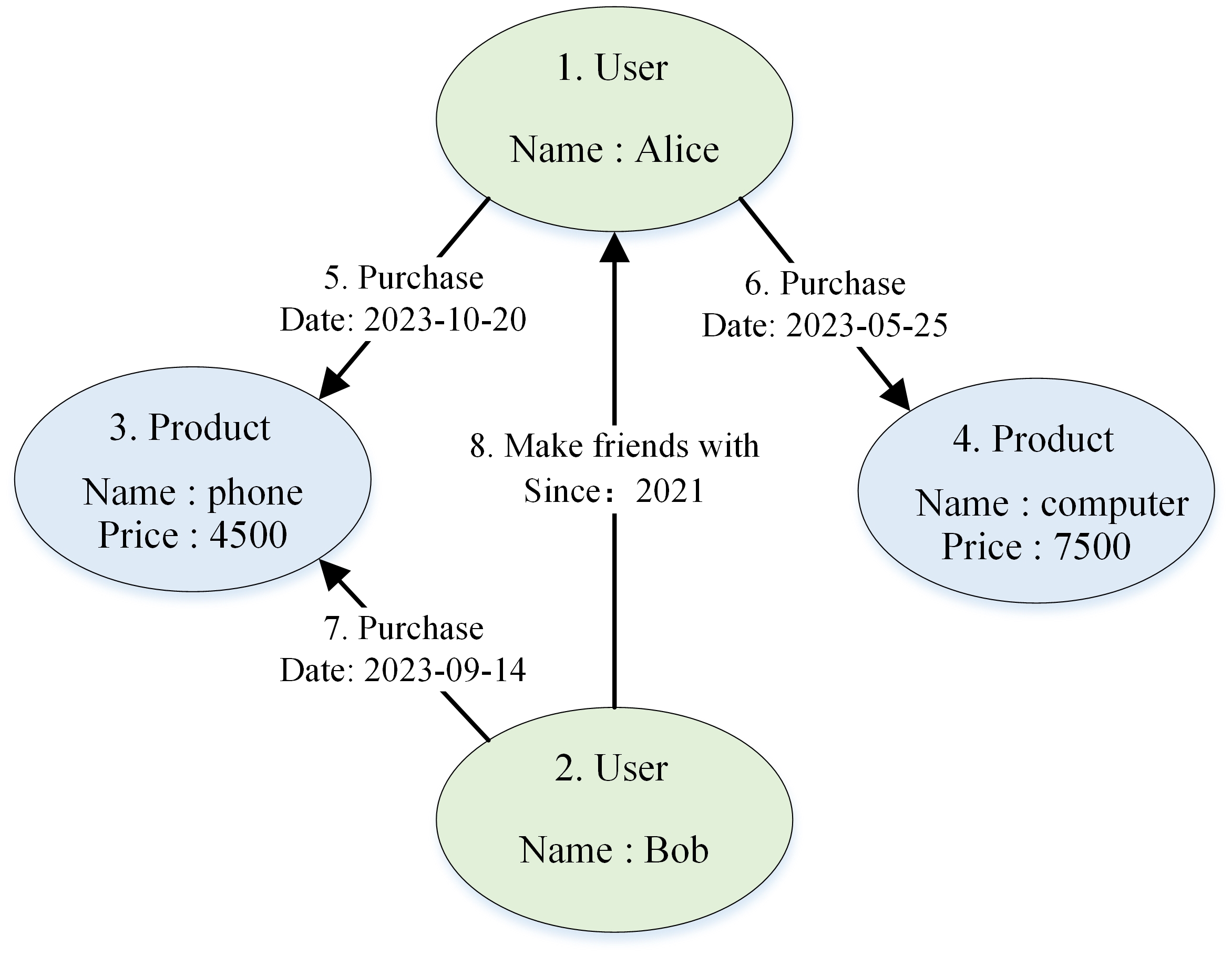}
		\caption{An example of labeled property graph}
	\end{figure}
	
	\section{A Simple Paradigm for Detecting Graph Database Engine Bugs}

	In this section, we introduced a simple paradigm, DGDB, that utilizes a large language model for automatically detecting bugs in graph database engines. As depicted in Figure 2, this paradigm mainly consists of three stages. In the first stage, graph data used for testing bugs is generated based on randomly generated graph schema. In the second stage, the graph data, instructions for generating queries, and constraints for query generation are input into a large language model (such as ChatGPT) to generate queries for testing bugs. In the third stage, after inputting generated queries into different graph database engines, detecting the presence of bugs is based on the comparison of consistency of the query results.  We choose the most popular ChatGPT as the large language model used in the paradigm.

	\subsection{GRAPH DATA \uppercase{Generation}}
	Graph databases utilize various graph models to store graph data, with the most commonly used being the labeled property graph model\cite{27} and the Resource Description Framework (RDF) graph model\cite{28}.

	In the Labeled Property Graph Model, both nodes and edges in the graph are endowed with labels and properties. Labels represent the type of nodes or edges, while properties describe the characteristics of nodes or edges in the form of key-value pairs. As shown in Figure 1, this labeled property graph contains two nodes labeled as "user", two nodes labeled as "product", three edges labeled as "purchase", and one edge labeled as "make friends with". The numbers to the left of the labels represent IDs. Nodes labeled as "user" have a "name" property, while nodes labeled as "product" have "name" and "price" properties. Edges labeled as "purchase" have a "date" property representing the purchase date, and edges labeled as "make friends with" have a "since" property indicating the date when the friendship with the user was established. The Labeled Property Graph Model provides a flexible and intuitive alternative for data modeling, precisely capturing and expressing the intricate relationships between nodes.
	
	The RDF graph model utilizes triples to describe relationships between resources, with each triple consisting of a subject, predicate, and object. The subject represents a resource, typically depicted as a node or edge in graph data. The object signifies information or values related to the subject, often represented as nodes, edges, or actual values in graph data. The predicate denotes the relationship between the subject and object. For example, the RDF triple describing node 1 in Figure 1 can be represented as: (subject: node 1, predicate: type, object: User), (subject: node 1, predicate: name, object: Alice). The RDF graph model benefits from its triple structure, providing stronger semantic expressiveness, and is widely applied in the field of knowledge graphs. However, when dealing with large-scale graph data, using triples requires more storage space and complex indexing structures, resulting in decreased query efficiency.  Therefore, we opt for the labeled property graph model to generate a graph database for testing bugs. Specifically, we randomly generate the graph schema of the labeled property graph and, based on this schema, randomly generate graph data in the graph database engine.

	\textbf{Graph schema generation.} The graph schemas generated in the labeled property graph should include node label set $NodeLabel$
	$Set$, edge label set $EdgeLabelSet$, and property set $PropertySet$. where $NodeTypeSet$, $EdgeLabelSet$, and $PropertySet$ represent the sets of node labels, edge labels, and properties of nodes and edges that appear in the graph, respectively. In our experiment, we let $NodeLabelSet = \left\{ {n{t_0},n{t_1},n{t_2},n{t_3}} \right\}$, $EdgeLabelSet = \, $\{ ${e{t_0},e{t_1},e{t_2}},$ ${e{t_3}}$\}, $PropertySet = \left\{ {name,{p_0},{p_1},...,{p_9}} \right\}$. In order to ensure diversity in graph data within the graph database engine, we use $name$ within $PropertySet$ to represent the unique name of nodes or edges, while the remaining property keys correspond to different types of property values, including $Integer$, $Float$, $String$, and $Boolean$. Additionally, a node or edge can simultaneously contain multiple different property key-value pairs.
	
	\textbf{Graph data generation.} Based on the above graph schema, we generate graph data stored in the graph database engine. The algorithm for generating graph data is shown as Algorithm 1 and consists of two parts. In the first part (Lines 1-7), for any generated node $v$, we first randomly select any label from $NodeLabelSet$ as the label for that node. Then, based on the $p$ distribution, we select single or multiple properties from $PropertySet$ as the properties for that node and generate property values of consistent types randomly. In the second part (Lines 8-15), we first randomly select two different nodes from $v\_set$ as the in-node and out-node of an edge. We then randomly select any label from $EdgeLabelSet$ as the label for that edge. Similarly, based on the $p$ distribution, we select single or multiple properties from $PropertySet$ as the properties for that edge and generate property values of consistent types randomly. We set the $p$ distribution to [0.8, 0.1, 0.05, 0.05], and the corresponding numbers of property key-value pairs for nodes or edges are [1, 2, 3, 4]. Note that the $name$ property key-value pairs are not included in this count.
	
	\begin{algorithm}[tbp]
		\renewcommand{\algorithmicrequire}{\textbf{Input:}}
		\renewcommand{\algorithmicensure}{\textbf{Output:}}
		\caption{Graph data generation}
		\label{alg:1}
		\begin{algorithmic}[1]
			\REQUIRE $NodeLabelSet$, $EdgelabelSet$, $PropertySet$, the number of nodes $M$, the number of edges $N$, the property key-value pair generation distribution $p$.
			\STATE $v\_set \leftarrow \emptyset $, $e\_set \leftarrow \emptyset $
			\FOR{$i = 1, 2,…,M$}
			\STATE $v \leftarrow $ create a new node.
			\STATE $v.label \leftarrow random.choice\left( {VerLabelSet} \right)$
			\STATE $v.properties \leftarrow $ generate the properties according to $p$ and $PropertySet$.
			\STATE $v\_set \leftarrow v\_set \cup v$
			\ENDFOR
			\FOR{$j = 1, 2,…,N$}
			\STATE ${v_1} \leftarrow random.choice\left( {v\_set} \right)$.
			\STATE ${v_2} \leftarrow random.choice\left( {v\_set - {v_1}} \right)$
			\STATE $e \leftarrow $ create a new edge with $v_1$ as the incoming node and $v_2$ as the outgoing node.
			\STATE $e.label \leftarrow random.choice\left( {EdgeLabelSet} \right)$.
			\STATE $e.properties \leftarrow $ generate the properties according to $p$ and $PropertySet$.
			\STATE $e\_set \leftarrow e\_set \cup e$.
			\ENDFOR
		\end{algorithmic}  
	\end{algorithm}
	
	\begin{figure*}[htbp]
		\centering
		\includegraphics[width=\linewidth]{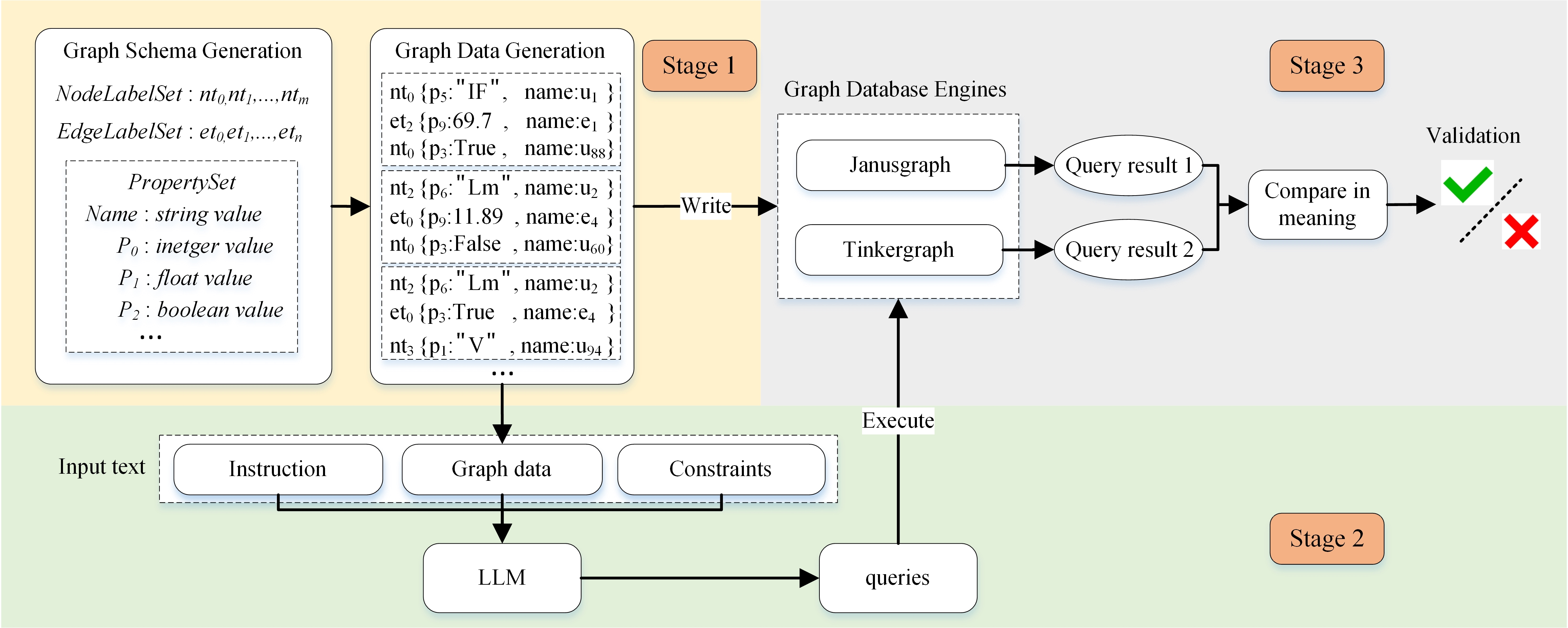}
		\caption{Overview of DGDB}
	\end{figure*}

	\subsection{\uppercase{Query Generation}}
	Traditional methods for generating queries\cite{17,18,19} often require designing complex algorithms and an extensive prior knowledge to achieve a high non-empty-result query ratio. In contrast, ChatGPT simplifies this process. With its demonstrated ability in zero-shot in-context learning across various text generation tasks, ChatGPT can achieve high non-empty-result query ratio by simply receiving the appropriate instructions alone\cite{29,30,31}.
	
	\hspace{-0.45cm} \rule{8.5cm}{1.0pt}
	
	\vspace{0.1cm}
	
	\textbf{Instruction: Your task is to generate queries in the Graph database according to the nodes and relationships in the mentioned graph. The edges in the graph are represented as (node type {attribute key value pairs})-(relationship type {attribute key value pairs})->(node type {attribute key value pairs}). For example, (nt0 {p5: "IF",name: "u1"})-(et2 {name: "e1",p9: 69.7}]->(nt0 {p3: true,name: "u88"}) indicates that there exists a directed edge of type et2 from a node named "u1" to a node named "u88".}
	
	\vspace{0.25cm}
	\textbf{The following is the graph data in the Graph database engine, which contains 100 nodes and 200 edges:}
	
	\textbf{(nt0 {p5: "IF",name: "u1"})-(et2 {name: "e1",p9: 69.7})->(nt0 {p3: true,name: "u88"}).}
	
	\textbf{(nt0 {p5: "IF",name: "u1"})-(et0 {p6: "7W",name: "e0"})->(nt0 {p1: "a",name: "u21"}).}

	\textbf{.  .  .}
	
	\textbf{(nt1 {p3: false,name: "u99"})-(et0 {p0: 69.37,name: "e199"})->(:nt0 {p0: 64.66,p1: "gL",name: "u30",p8: 51.93}).}
	
	\vspace{0.25cm}
	\textbf{Query: Based on the instruction and graph data, Please generate generate a specific number of queries with the different operators (eg. some operators used in a specific graph query language.) and meet the following conditions: (Some constraints.)}
	
		\vspace{0.1cm}
	
	\hspace{-0.45cm} \rule{8.5cm}{1.0pt}
	
	To utilize ChatGPT for generating meaningful queries, we start by giving ChatGPT a specific task through Instructions. Next, we explain the specific constructions of graph data to it using examples. This step is crucial to offer ChatGPT an unbiased perception of the graph data. Then we input the graph data into ChatGPT, based on which allows it to understand the complex structure and semantic relationship of graph data. Finally, we ask ChatGPT to generate corresponding query statements. To enhance the diversity of queries and improve the ratio of non-empty -result query, we set various constraints in the query formulation process.
	
	\vspace{0.25cm}
	\hspace{-0.35cm}\textbf{Listing 1: An example of a query that is consistent in meaning but not entirely consistent in return results.}
	\vspace{0.25cm}
	
	\hspace{-0.39cm}
	\fbox{%
		\parbox{8.3cm}{
			1. \textbf{Query}: \textbf{\textcolor[RGB]{0, 128, 0}{MATCH}} (n) \textbf{\textcolor[RGB]{0, 128, 0}{WHERE}} n.name = 'u16' \textbf{\textcolor[RGB]{0, 128, 0}{RETURN}} n; 
			
			\vspace{0.05cm}
			
			2. \textbf{Neo4j}: [Node('nt0', name='u16', p4=False, p7=True, p9=41.96)]. 
			
			\vspace{0.05cm}
			
			3. \textbf{Agensgraph}: ('nt0[4.4]\{"p4": false, "p7": true, "p9": 41.96, "name": "u16"\}',).
		}%
	}

	\subsection{\uppercase{Bug Detection}}
	
	The differential testing method\cite{32}, is widely used to detect bugs and errors in software programs  due to its ability to discover new problems, facilitate continuous improvements, and be applied universally. Therefore, we use the differential testing method to detect bugs in the graph database engine. Specifically, our process begins by selecting graph databse queries generated by ChatGPT as initial test samples. Subsequently, we refine selected queries by filtering out those that either yield random query results or alter the original graph data to form our final set of test samples. Following that, we execute the same test samples on different graph database engines. Ultimately, we compare the outputs of different graph database engines. Since different graph database engines handle nodes or edges differently, requiring completely consistent query results may lead to discrepancies that not necessary reflect the existence of bugs. As shown in listing 1, the query results are not completely consistent, but they all represent node 'u16'. Therefore, in order to improve the efficiency of bug detection, we believe that if the query results of different database engines are consistent in meaning, then the test sample passes; otherwise, a bug may have occurred.

	\section{DGDB Examples}
	
	In this section, we provide two examples demonstrating how to use DGDB to detect bugs in graph database engines: DGDB-Cypher for detecting bugs in graph database engines using Cypher query language and DGDB-Gremlin for detecting bugs in graph database engines using Gremlin query language.
	
	\subsection{\uppercase{DGDB-Cypher}}
	For the bug detection in graph database engines based on the Cypher query language, we selected Neo4j and AgensGraph as the target databases. Neo4j\cite{6} is a high-performance open-source graph database engine designed for storing, querying, and analyzing graph data. It is widely used due to its efficient execution of complex graph queries and the capability to handle large-scale graph data. AgensGraph\cite{11} is a multi-model graph database engine known for its flexibility in handling different types of data, including graph data and relational data, making it widely utilized.
	
	DGDB-Cypher first constructs corresponding graph data in Neo4j and AgensGraph, ensuring that the node types, node properties, edge types and edge properties are identical. The constructed graph data is then inserted into the instruction, fed into ChatGPT, and used to generate Cypher queries. The text inputted into ChatGPT is as follows:
	
	\hspace{-0.45cm} \rule{8.5cm}{1.0pt}
	
	\textbf{Instruction: Your task is to generate queries in the Graph database according to the nodes and relationships in the mentioned graph.} \textbf{The edges in the graph are represented as \(|\)  (: node type {attribute key value pairs})\(|\) (: relationship type {attribute key value pairs})\(|\) (: node type {attribute key value pairs})\(|\). For example, \(|\) (: nt8 {p6: false, name: "u97"}) \(|\)  (: et11 {p8: true}) \(|\)  (: nt6 {p17: false, name: "u33"}) \(|\) Indicates that there exists a directed edge of type et11 from a node named "u97" to a node named "u33".}
	
	\vspace{0.25cm}
	\textbf{The following is the graph data in the Graph database engine, \textcolor{gray}{which contains 100 nodes and 200 edges:}}
	
	\textbf{\textcolor{gray}{\(|\)(:nt0 {p5: "IF",name: "u1"})\(|\)(:et2 {name: "e1",p9: 69.7})\(|\)(:nt0 {p3: true,name: "u88"})\(|\)}}
	
	\textbf{\textcolor{gray}{\(|\)(:nt0 {p5: "IF",name: "u1"})\(|\)(:et0 {p6: "7W",name: "e0"})\(|\)(:nt0 {p1: "a",name: "u21"})\(|\)}}
	
	\textbf{\textcolor{gray}{\(|\)(:nt2 {p6: "Lm",name: "u2"})\(|\)(:et0 {p1: "Glo",name: "e4",p9: 11.89})\(|\)(:nt0 {p3: false,name: "u60"})\(|\)}}
	
	\textbf{. . .}
	
	\textbf{\textcolor{gray}{\(|\)(:nt1 {p3: false,name: "u99"})\(|\)(:et0 {p0: 69.37,name: "e199"})\(|\)(:nt0 {p0: 64.66,p1: "gL",name: "u30",p8: 51.93})\(|\)}}
	
	\vspace{0.25cm}
	\textbf{Based on the instruction and graph database, Please generate} \textbf{\textcolor{gray}{twenty cypher queries with the different operators(eg. MATCH, OPTIONAL MATCH, WHERE, Aggregation, FOREACH, RETURN, ORDER BY, WITH, UNWIND, UNION, UNION ALL, collect, predicate, coalesce, length, type, keys, labels, startNode, endNode, nodes, relationships, reduce, shortestPath)} and meet the following conditions:}
	
	\textbf{1. Please make sure the queried data is the node or link mentioned earlier.}
	
	\textbf{2. Please ensure that the values in the attribute key value pairs of the constraint exist.}
	
	\textbf{3. If you want to generate a query with relationships, please pay attention to the direction of the query relationships in the generated query statement.}
	
	\textbf{4. Please ensure that the generated queries use different keywords as much as possible.}
	
	\textbf{5. Please ensure that the generated query statement will not change the data of the Graph database\textcolor{gray}{(eg. Do not use the create operators).}}

	\hspace{-0.45cm} \rule{8.5cm}{1.0pt}
	
	In the above text, the modified parts are highlighted in gray. It can be observed that we mainly made changes to the types, quantity, operators and constraints of generated queries.
	
	\begin{itemize}[leftmargin=*]
		
		\item \textbf{Types and quantity:} We stipulate the generation of Cypher queries; and due to the token limit of ChatGPT, we set the quantity of generated Cypher queries to 20.
		
		\item \textbf{Operators:} We hope that ChatGPT uses different operators when generating queries to avoid situations where the generated queries are highly similar, as shown in Listing 2, and to improve detection efficiency.
		
		\item \textbf{Constraints:} We have set a total of five constraints. The first two points indicate that we hope ChatGPT can generate as many non-empty-result queries as possible. The third constraint is to ensure that ChatGPT considers the direction of edges when generating query statements because we found that without this constraint, ChatGPT tends to ignore the direction of arrows in most generated query statements. The fourth constraint is to ensure that ChatGPT can generate more complex query statements. The fifth constraint is to ensure consistency in graph data across different graph database engines.
	\end{itemize}
	
	\vspace{0.25cm}
	\hspace{-0.35cm}\textbf{Listing 2: An example of generating highly similar queries, where only the types of nodes or edges that they belong to are changed in the queries.}
	\vspace{0.25cm}
	
	\hspace{-0.39cm}
	\fbox{%
		\parbox{8.3cm}{
			1. \textbf{\textcolor[RGB]{0, 128, 0}{MATCH}} (n:nt1)-[:et1]->(m:nt3) \textbf{\textcolor[RGB]{0, 128, 0}{RETURN}} n, m;
			
			\vspace{0.05cm}
			
			2. \textbf{\textcolor[RGB]{0, 128, 0}{MATCH}} (n:nt1)-[:et1]->(m:nt2) \textbf{\textcolor[RGB]{0, 128, 0}{RETURN}} n, m; 
			
		}%
	}
	\vspace{0.25cm}
	
	The above modifications were mostly made to generate high-quality Cypher query statements, improving the efficiency of detecting bugs. We iterated a total of 200 times, generating 4000 query statements. We then filtered out queries containing operators that produce random results, such as queries including the skip or limit keywords. The remaining query statements were executed on Neo4j and AgensGraph graph database engines, and the query results were compared. Due to discrepancies in internal implementation, data storage structure, and performance optimization among different graph database engines, there may be variations in query results. As shown in Listing 3, the returned results consist of lists containing nodes such as $u1$, $u3$, but their presentation may differ. Compared to using a mapping table for data transformation, which introduces some additional computational and storage costs\cite{18}, our method utilizes regular expressions\cite{42} to directly extract information about corresponding nodes from query results, enhancing the efficiency of bug detection. Finally, we compare this information, and if discrepancies arise, it indicates that the graph database engine may have a bug in executing this query statement.

	\vspace{0.25cm}
	\hspace{-0.35cm}\textbf{Listing 3: An example of using regular expressions to extract information from cypher query results.}
	\vspace{0.25cm}
	
	\hspace{-0.39cm}
	\fbox{%
		\parbox{8.3cm}{
			1. \textbf{Query}: \textbf{\textcolor[RGB]{0, 128, 0}{MATCH}} (n:nt0)  \textbf{\textcolor[RGB]{0, 128, 0}{WITH collect}}(n) as n \textbf{\textcolor[RGB]{0, 128, 0}{RETURN DISTINCT}} n; 
			
			\vspace{0.05cm}
			
			2. \textbf{Neo4j}: [Node('nt0', name='u1', p5='IF'), Node('nt0', name='u3', p4=True, p6='9')...] - regular expression -> [\{name : 'u1', p5 : 'IF'\}, \{name : 'u3', p4 : True, p6 : '9'\}...]. 
			
			\vspace{0.05cm}
			
			3. \textbf{Agensgraph}: [\{'id': '4.1', 'tid': '(0,1)', 'properties': \{'p5': 'IF', 'name': 'u1'\}\}, \{'id': '4.2', 'tid': '(0,2)', 'properties': \{'p4': True, 'p6': '9', 'name': 'u3'\}\}...] - regular expression -> [\{name : 'u1', p5 : 'IF'\}, \{name : 'u3', p4 : True, p6 : '9'\}...].
		}%
	}

	\subsection{\uppercase{DGDB-Gremlin}}
	
	For the bug detection in graph database engines based on the Gremlin query language, we selected JanusGraph and TinkerGraph as the target database engines. JanusGraph\cite{7} is a high-performance distributed graph database and widely used for handling large-scale graph data and complex graph queries. TinkerGraph\cite{10} is a lightweight database in the TinkerPop graph computing framework, widely applied in prototype development for small-scale projects due to its superior read and write performance storing data in memory.
	
	DGDB-gremlin first constructs graph data in JanusGraph and TinkerGraph separately, ensuring consistency in the stored graph data between the two graph database engines. The constructed graph data is then converted into the context, which is then fed into ChatGPT to generate Gremlin query statements. The specific text input to ChatGPT is as follows:
	
	\hspace{-0.45cm} \rule{8.5cm}{1.0pt}
	
	\textbf{Instruction: Your task is to generate queries in the Graph database according to the nodes and relationships in the mentioned graph.} \textbf{The edges in the graph are represented as (node type {attribute key value pairs})-(relationship type {attribute key value pairs})->(node type {attribute key value pairs}). For example, (nt0 {p5: "IF",name: "u1"})-(et2 {name: "e1",p9: 69.7}]->(nt0 {p3: true,name: "u88"}) Indicates that there exists a directed edge of type et2 from a node named "u1" to a node named "u88".}
	
	\vspace{0.25cm}
	\textbf{The following is the graph data in the Graph database engine, \textcolor{gray}{which contains 100 nodes and 200 edges:}}
	
	\textbf{\textcolor{gray}{(nt0 \{p5: "IF",name: "u1"\})-(et2 \{name: "e1",p9: 69.7\}]->(nt0 \{p3: true,name: "u88"\})}}

	\textbf{.  .  .}
	
	\textbf{\textcolor{gray}{(nt1 \{p3: false,name: "u99"\})-(et0 \{p0: 69.37,name: "e199"\})->(:nt0 \{p0: 64.66,p1: "gL",name: "u30",p8: 51.93\}).}}

	\vspace{0.25cm}
	\textbf{Based on the instruction and graph database, Please generate} \textbf{\textcolor{gray}{twenty gremlin queries with the different operators(eg. hasLabel(), hasId(), has(), hasNot(), values(), label(), id(), properties(), values(), valueMap(), select(), dedup(), local(), order().
	by(), where(),filter(),match(),eq(), neq(), gt(), gte(), inside(), outside(), group().by(), groupCount().by(), in(), out(), inE(), outE(), inV(), outV(), both(), path(), repeat().until(), sum(), max(), min(), mean(),contains(), choose(), union(), fold())} and meet the following conditions:}
	
	\textbf{1. Please make sure the queried data is the node or link mentioned earlier.}
	
	\textbf{2. Please ensure that the values in the attribute key value pairs of the constraint exist.}
	
	\textbf{3. If you want to generate a query with relationships, please pay attention to the direction of the query relationships in the generated query statement.}
	
	\textbf{4. Please ensure that the generated queries use different keywords as much as possible.}
	
	\textbf{5. Please ensure that the generated query statement will not change the data of the Graph database\textcolor{gray}{(eg. Do not use the addV() or addE() operators).}}

	\hspace{-0.45cm} \rule{8.5cm}{1.0pt}
	
	Similar to the Instruction for generating Cypher query statements, we made modifications to the types, quantity, operators, and constraints for generating Gremlin query statements. The details are as follows: 
	
	\begin{itemize}[leftmargin=*]
		
		\item \textbf{Types and quantity:} We set the total 20 types of generated query statements to Gremlin.
		
		\item \textbf{Operators:} We select a large number of commonly used operators in Gremlin to ensure the diversity of the generated Gremlin query statements.
		
		\item \textbf{Constraints:} The constraints for generating Gremlin query statements are generally similar to those for generating Cypher query statements. Both aim for ensuring complex structures and high non-empty-result ratio for queries generated by ChatGPT, thereby improving the efficiency of detecting graph database engine bugs.
		
	\end{itemize}
	
	Like DGDB-CYPHER, we use the differential testing method to detect bugs for Gremlin-based graph database engines. Notably,  both JanusGraph and TinkerGraph return results as node ID for node querying requests. Furthermore, since JanusGragh automatically generate node ID within the graph database for maximizing data consistency, it is challenging for us to directly determine whether the query results are identical solely based on node IDs. Our solution is to compare the query results by obtaining the correspondent properties to the node IDs, as exhibited in Listing 4.
	
	\vspace{0.25cm}
	\hspace{-0.35cm}\textbf{Listing 4: An example of comparing node IDs in gremlin query results.}
	\vspace{0.25cm}
	
	\hspace{-0.39cm}
	\fbox{%
		\parbox{8.3cm}{
			1. \textbf{Query}: g.\textcolor{orange}{V}().\textcolor{orange}{hasLabel}('nt0'); 
			
			\vspace{0.05cm}
			
			2. \textbf{Janusgraph}: [v[16416], v[24608], v[32800],...]. 
			
			\vspace{0.05cm}
			
			3. \textbf{Tinkergraph}: [v[3], v[9], v[278],...].
			
			\vspace{0.05cm}
			
			4. \textbf{Query}: g.\textcolor{orange}{V}().\textcolor{orange}{hasLabel}('nt0').\textcolor{orange}{valueMap}()
			
			\vspace{0.05cm}
			
			5. \textbf{Janusgraph}: [\{'name': ['u16'], 'p9': [41.96], 'p4': [False], 'p7': [True]\}, \{'name': ['u25'], 'p5': ['UR']\},...]
			
			\vspace{0.05cm}
			
			6. \textbf{Tinkergraph}: [\{'p5': ['IF'], 'name': ['u1']\}, \{'p4': [True], 'p6': ['9'], 'name': ['u3']\},...]
			
		}%
	}
	
	\section{\uppercase{Evaluations}}
	
	To demonstrate the effectiveness of the proposed DGDB paradigm, we compare DGDB-cypher and DGDB-gremlin respectively with the GDSmith and Grand methods, and address the following research questions:
	\begin{itemize}[leftmargin=*]
		\item(RQ1) How does the quality of queries generated by ChatGPT stand in comparison to those produced by the baseline model?
		\item(RQ2) Can the graph database engine bug detection methods proposed based on the DGDB paradigm detect real-world bugs in popular graph database engines?
	\end{itemize}
	
	\subsection{\uppercase{Evaluation Setup}}
	
	\subsubsection{Subjects}
	
	We use listed graph database engines in Table 2 for detecting bugs, and select the latest versions in all cases since new versions of graph database engines may have addressed bugs presented in previous versions while detecting bugs in previous versions might reveal issues that have already been fixed. Moreover, detecting bugs in new versions that are more likely to be persisted in old versions makes our tasks more meaningful.
	
	\begin{table}[htb]
		\caption{The graph database engines selected for testing}
		\begin{tabular}{cccc}
			\toprule
			GDBMS       & Version & Release date & Supported languages \\ \midrule
			Neo4j       & 4.4.26  & 2023.9.20    & Cypher              \\ 
			Agensgraph  & 2.13.0  & 2022.10.13   & Cypher              \\ 
			Janusgraph  & 1.0.0   & 2023.10.21   & Gremlin             \\ 
			Tinkergraph & 3.7.0   & 2023.7.31    & Gremlin             \\ \bottomrule
		\end{tabular}
	\end{table}
	
	\subsubsection{Baseline}
	To demonstrate the superiority of the proposed paradigm, we compare two methods for graph database engine bug detection, as follows:
	\begin{itemize}[leftmargin=*]
		
		\item GDSmith\cite{19}: GDSmith is a bug detection method for Cypher-based graph database engines. It first guides the generation of semantically valid Cypher queries based on property graphs. Subsequently, the generated queries are executed on each graph database engine and identify potential bugs     based on discrepancies of the outputs.

		\item Grand\cite{18}: Grand is a bug detection method for Gremlin-based graph database engines. It first constructs a traversal model based on Gremlin API, based on which to generate effective Gremlin queries. Subsequently, the generated queries are executed on different graph database engine, and bugs are detected based on discrepancies between the query results.
		
	\end{itemize}
	
	\subsubsection{Implementation details}
	
	In our research, we implement DGDB-cypher and DGDB-gremlin through  Python 3.8. We use the Py2neo package to connect to the Neo4j for operations, the Psycopg2 package to connect to the AgensGraph for operations, and the Gremlin-python package to connect to JanusGraph and TinkerGraph for operations. The version of the ChatGPT model is gpt-3.5-turbo-16k-0613. The versions of each package are as shown in Table 3. The numbers of generated nodes, edges, node types, edge types, and property types are set to 100, 200, 4, 4, and 11, respectively. We set the number of rounds for both DGDB-cypher and DGDB-gremlin to 200, generating 4000 query statements respectively.
	
	\begin{table}[htb]
		\caption{The version of each package we used}
		\begin{tabular}{cc}
			\toprule
			Package        & Version  \\ \midrule
			Py2neo         & 2021.2.3 \\
			Psycopg2       & 2.9.7    \\
			Gremlin-python & 3.2.6    \\
			Openai         & 0.27.10  \\ \bottomrule
		\end{tabular}
	\end{table}
	
	\subsection{\uppercase{Performance Comparison (RQ1)}}
	
	Grand\cite{18}, as the first method proposed for detecting graph database engine bugs, utilizes the differential testing method to detect bugs in graph database engines using the gremlin query language. The method ran for 2400 seconds, generating a total of 15,000 queries. It  successfully detects a total of 21 bugs across six graph database engines, with most of the bugs already being fixed. However, this method often detects a significant number of discrepancies, as evidenced by the 709 discrepancies found among 15,000 generated queries in the original study and the 615 discrepancies observed in 10,000 generated queries when Grand is used to detect bugs in Janusgraph, TinkerGraph and HugeGraph, as mentioned in \cite{17}. These discrepancies are largely attributed to variations in handing exceptions by different graph database engines rather than detecting actual bugs in graph database engines \footnote{https://github.com/choeoe/Grand/issues/1}. GDSmith\cite{19} is the first graph database bug detection method based on the Cypher query language. It ran for 12 hours in neo4j, redisgraph, and memgraph, discovering a total of 11,275 discrepancies and identifying 28 bugs from these discrepancies, with the majority of them already fixed. It can be observed that both methods mentioned above have discovered a significant number of discrepancies. However, a large number of discovered discrepancies may merely indicate the presence of numerous duplicated bugs or results from different internal implementations of various graph database engines, leading to different outputs. Furthermore, analyzing each discrepancy can result into substantial time cost, and naturally compromise the efficiency of bug detection. Therefore, we believe that the quantity of discrepancies should not be used as a benchmark for the performance of graph database engine bug detection. Moreover, since most of the bugs detected by the two methods mentioned above have already been fixed through the version updates, we cannot simply compare the performance of graph database engine bug detection methods based solely on the number of detected bugs. Our methods DGDB-cypher and DGDB-gremlin each ran for about 4200 seconds, generated 4000 queries, and collectively detected 10 unfixed bugs in the four latest versions of popular graph database engines, which exclude the solved bugs, and therefore, offer more convincing evidences for demonstrating the effectiveness of our methods.
	
	DGDB-cypher, DGDB-gremlin, Grand, and GDSmith all use the differential testing method to detect graph database engine bugs, where the quality of generated queries determines the performance of differential testing methods. Therefore, we analyzed the quality of queries generated by these four methods. We believe that high-quality queries should possess two essential criteria: it should be diversified and have a high non-empty-result query ratio. Specifically, diverse queries can comprehensively detect potential graph database bugs introduced by different operators, and high non-empty-result queries can significantly improve the efficiency of bug detection.
	
	\vspace{0.25cm}
	\hspace{-0.35cm}\textbf{Listing 5: An example of a query on search paths.}
	\vspace{0.25cm}
	
	\hspace{-0.39cm}
	\fbox{%
		\parbox{8.3cm}{
			1. \textbf{\textcolor[RGB]{0, 128, 0}{MATCH}} p=\textbf{\textcolor[RGB]{0, 128, 0}{shortestPath}}((n1)-[:et0*]-(n2:nt0)) \textbf{\textcolor[RGB]{0, 128, 0}{WHERE}} n1 <> n2 \textbf{\textcolor[RGB]{0, 128, 0}{RETURN}} n1, n2;
		}%
	}
	\vspace{0.05cm}
	
	\begin{itemize}[leftmargin=*]
		
		\item \textbf{Diversity}. The Grand method constructs a traversal model to generate queries, including three types of operations: filter, predicate, and aggregate. However, due to the limitations of this traversal model, the generated queries may not include certain operators such as $repeat()$, $until()$, $path()$, $fold()$, etc., which hinder the detection of bugs associated with these operators. The GDSmith method firstly constructs a Cypher skeleton, then generates subgraphs that match the property graph to complete the pattern of the Cypher skeleton. Finally, based on static query analysis, the conditional expressions of the Cypher skeleton are completed to obtain the generated queries. However, this method only queries elements in the matching subgraphs and cannot generate statements for searching paths, as shown in Listing 5. It can be seen that, due to limitations in the methods of generating queries, Grand and GDSmith cannot generate more comprehensive queries. If one wants to increase the diversity of generated queries, extensive expert knowledge is required to make complex modifications to existing methods, resulting in excessive manual cost. Comparably, DGDB-cypher and DGDB-gremlin exhibit advantages in utilizing more flexible and convenient operators in the form of textual input to ChatGPT. For example, in DGDB-gremlin, "\textcolor{gray}{\textbf{Please generate twenty gremlin queries with different types of methods (e.g., hasLabel(), hasId(), has(), hasNot(), values(), [operators])}}", we only need to replace the \textcolor{gray}{\textbf{[operators]}} with the operators we want to use in the generated queries to ensure that the generated queries include that operator, creating diverse queries for a more comprehensive detection of bugs in the graph database engine.
		
		\item \textbf{Non-empty-result query ratio}. As shown in Table 4, we compared the non-empty-result query ratios generated by DGDB-cypher, DGDB-gremlin, Grand, and GDSmith. It can be observed that the non-empty-result query ratio of queries generated by DGDB-cypher and DGDB-gremlin is higher than those by GDSmith and Grand. This indicates that ChatGPT can understand graph data well and generate high-quality queries based on constraints. Additionally, the non-empty-result query ratio of Grand is much lower than GDSmith. This is because Grand is a random differential testing method, while GDSmith improves the non-empty-result query ratio by using graph-guided pattern generation and data-guided condition generation.
		
	\end{itemize}
	
	In summary, compared to existing methods for detecting bugs in graph database engines, utilizing the DGDB paradigm allows for a more comprehensive and efficient detection of bugs in graph database engines.
	
	\begin{table}[htb]
		\caption{ The non-empty-result query ratio of DGDB and
			two baselines}
		\begin{tabular}{ccc}
			\toprule
			Approach     & Language & Non-empty-result Query Ratio \\ \midrule
			DGDB-cypher  & Cypher   & 79.06\%                      \\
			GDSmith      & Cypher   & 73.66\%                      \\
			DGDB-gremlin & Gremlin  & 80.33\%                      \\
			Grand        & Gremlin  & 40.30\%                      \\ \bottomrule
		\end{tabular}
	\end{table}
	
	\subsection{\uppercase{Detected Wrong-Result Bugs (RQ2)}}
	
	Bugs detected using differential testing mainly fall into two categories: crash bugs\cite{45} and wrong-result bugs\cite{19}. Crash bugs occur when users run certain commands and do not receive the expected results; instead, it causes the graph database engine to crash. Since users cannot obtain the expected results, they become aware of the bug. Grand and GDSmith detect many bugs of this type. Wrong-result bugs are refereed to the incorrect results obtained based on users' running commands. These erroneous results often resemble with correct ones, making it challenging for users to identify manually and, consequently, leading to greater risks compared to the crash bugs. Therefore, our primary focus is on detecting wrong-result bugs.
	
	\begin{table}[htb]
		\caption{Bugs that we found in the tested graph database engines}
		\begin{tabular}{ccc}
			\toprule
			Approach     & \makecell{Graph database\\engine} & \makecell{Number of wrong-result\\ bugs detected} \\ \midrule
			DGDB-cypher  & Neo4j   & 2                     \\
			DGDB-cypher  & Agensgraph   & 5                   \\
			DGDB-gremlin & Janusgraph  & 3                   \\
			DGDB-gremlin & Tinkergraph  & 0                     \\ \bottomrule
		\end{tabular}
	\end{table}

	Table 5 indicates the number of bugs detected using DGDB-cypher and DGDB-gremlin, and we present the specific bugs in the subsequent content. Notably, we choose to maximize the clarity for our readers by presenting the succinct yet illustrative examples in the following content instead of the original queries that are implemented to discover the bugs since the original queries are typically too complex to be faithfully showcased in this paper.
	
	\vspace{0.25cm}
	\hspace{-0.35cm}\textbf{Listing 6: Cypher queries that trigger a wrong-result bug in Neo4j 4.4.26.}

	\vspace{0.25cm}
	
	\hspace{-0.39cm}
	\fbox{%
		\parbox{8.3cm}{
			\vspace{0.05cm}
			1. \textbf{Query 1:} \textbf{\textcolor[RGB]{0, 128, 0}{MATCH}} (n:nt3) \textbf{\textcolor[RGB]{0, 128, 0}{RETURN}} \textbf{\textcolor[RGB]{0, 128, 0}{count}}(n), \textbf{\textcolor[RGB]{0, 128, 0}{avg}}(n.p8);
			
			\vspace{0.05cm}
			
			2. \textbf{Neo4j:} [3, 25.78666666666667] \ding{56}
			\vspace{0.05cm}
			
			3. \textbf{Agensgraph:} (24, 25.786666666666665) \ding{52}
			\vspace{0.05cm}
			
			4. \textbf{Query 2:} \textbf{\textcolor[RGB]{0, 128, 0}{MATCH}} (n:nt3) \textbf{\textcolor[RGB]{0, 128, 0}{RETURN}} \textbf{\textcolor[RGB]{0, 128, 0}{count}}(n);
			
			\vspace{0.05cm}
			
			5. \textbf{Neo4j:} [26] \ding{52}
			\vspace{0.05cm}
			
			6. \textbf{Agensgraph:} (26,) \ding{52}
			\vspace{0.05cm}

		}%
	}
	
	\vspace{0.25cm}
	
	As shown in listing 6, query 1 returns the number of nodes of type "nt3" and the mean value of the "p8" property value for nodes of that type. It can be observed that Neo4j and AgensGraph return different node counts. However, the node count returned by query 2 is the same. Therefore, we believe the reason for the bug is that Neo4j, when simultaneously returning the node count and node property values, first considers whether the "p8" property value exists in the properties of nodes of type "nt3". If it exists, it is counted; otherwise, it is not counted, leading to the bug.

	\vspace{0.25cm}
	\hspace{-0.35cm}\textbf{Listing 7: Cypher queries that trigger a wrong-result bug in Neo4j 4.4.26.}
	\vspace{0.25cm}

	\hspace{-0.39cm}
	\fbox{%
		\parbox{8.3cm}{
			\vspace{0.05cm}
			1. \textbf{Query:} \textbf{\textcolor[RGB]{0, 128, 0}{MATCH}} (n:nt1)-[r]-() \textbf{\textcolor[RGB]{0, 128, 0}{RETURN}} n.name \textbf{\textcolor[RGB]{0, 128, 0}{AS}} Name, \textbf{\textcolor[RGB]{0, 128, 0}{sum}}(r.p8) \textbf{\textcolor[RGB]{0, 128, 0}{AS}} TotalP8;
			
			\vspace{0.05cm}
			
			2. \textbf{Neo4j:} [['u5', 0], ['u6', 17.01], ['u7', 17.02],...,['u22', 84.84]] \ding{56}
			\vspace{0.05cm}
			
			3. \textbf{Agensgraph:} [('u5', None), ('u6', 17.01), ('u7', 17.02),...,('u22', 84.84)] \ding{52}
			\vspace{0.05cm}
			
		}%
	}
	\vspace{0.25cm}
	
	As shown in listing 7, this query returns the names of nodes of type "nt1" and the sum of the "p8" property values for their respective edge sets. It can be observed that there is a difference between the sum of property values returned by Neo4j and AgensGraph. When AgensGraph returns with the "none" result, it indicates that the "p8" property does not exist in the edge set of the node. However, when Neo4j returns its result as "0", it might lead to the misconception that the sum of "p8" property values in the edge set of the node is 0, suggesting that "p8" exists in the edge set of the node. We believe this bug may be due to Neo4j's inadequate handling of exceptional values.

	\vspace{0.25cm}
	\hspace{-0.35cm}\textbf{Listing 8: Cypher queries that trigger a wrong-result bug in Agensgraph 2.13.0.}
	\vspace{0.25cm}

	\hspace{-0.39cm}
	\fbox{%
		\parbox{8.3cm}{
			\vspace{0.05cm}
			1. \textbf{Query 1:} \textbf{\textcolor[RGB]{0, 128, 0}{MATCH}} (n)-[r]->() \textbf{\textcolor[RGB]{0, 128, 0}{UNWIND}} n.p6 \textbf{\textcolor[RGB]{0, 128, 0}{AS}} values, \textbf{\textcolor[RGB]{0, 128, 0}{RETURN}} values;
			
			\vspace{0.05cm}
			
			2. \textbf{Neo4j:} [['Lm'], ['GOvy'], ['5Yz'],...,['Rk']] \ding{52}
			\vspace{0.05cm}
			
			3. \textbf{Agensgraph:} [] \ding{56}
			\vspace{0.05cm}
			
			4. \textbf{Query 2:} \textbf{\textcolor[RGB]{0, 128, 0}{MATCH}} (n)-[r]->() \textbf{\textcolor[RGB]{0, 128, 0}{RETURN}} n.p6;
			
			\vspace{0.05cm}
			
			5. \textbf{Neo4j:} [[None], ['Lm'], ['GOvy'],  [None],...,['Rk']] \ding{52}
			\vspace{0.05cm}
			
			6. \textbf{Agensgraph:} [(None,), ('Lm',), ('GOvy',), (None,),..., ('Rk',)] \ding{52}
			\vspace{0.05cm}
			
			7. \textbf{Query 3:} \textbf{\textcolor[RGB]{0, 128, 0}{MATCH}} (n)-[r]->() \textbf{\textcolor[RGB]{0, 128, 0}{WITH}} n.p6 \textbf{\textcolor[RGB]{0, 128, 0}{AS}} values, \textbf{\textcolor[RGB]{0, 128, 0}{RETURN}} values;
			
			\vspace{0.05cm}
			
			8. \textbf{Neo4j:} [[None], ['Lm'], ['GOvy'],  [None],...,['Rk']] \ding{52}
			\vspace{0.05cm}
			
			9. \textbf{Agensgraph:} [(None,), ('Lm',), ('GOvy',), (None,),..., ('Rk',)] \ding{52}
			\vspace{0.05cm}
			
		}%
	}
	\vspace{0.25cm}
	
	As shown in listing 8, query 1 unfolds the "p6" property values of nodes with outgoing relationships into separated rows. It can be observed that neo4j returns the correct results, while Agensgraph returns an empty set. Query 2 aims for checking if the data retrival functions properly by returning the "p6" property values of nodes with outgoing relationships, and the consistent results required from neo4j and Agensgraph confirm them being functional. Query 3 replaces the "unwind" keyword with "with" to check if other operators are causing the issue, and the results from neo4j and Agensgraph are also identical. Therefore, we suspect that there may be a bug in Agensgraph when using the "unwind" keyword to unfold property values.
	
	\vspace{0.25cm}
	\hspace{-0.35cm}\textbf{Listing 9: Cypher queries that trigger a wrong-result bug in Agensgraph 2.13.0.}
	\vspace{0.25cm}

	\hspace{-0.39cm}
	\fbox{%
		\parbox{8.3cm}{
			\vspace{0.05cm}
			1. \textbf{Query 1:} \textbf{\textcolor[RGB]{0, 128, 0}{MATCH}} (n:nt3 \{p5: 'Ce'\})-[:et3]->(m) \textbf{\textcolor[RGB]{0, 128, 0}{RETURN}} n, \textbf{\textcolor[RGB]{0, 128, 0}{COLLECT}}(m);
			
			\vspace{0.05cm}
			
			2. \textbf{Neo4j:} [Node('nt3', name='u4', p5='Ce'), [Node('nt1',name = 'u85', p3=True), Node('nt2', name='u84', p9=44.79)]] \ding{52}
			
			\vspace{0.05cm}
			
			3. \textbf{Agensgraph:} [('nt3[3.2]\{"p5": "Ce", "name": "u4"\}', [\{'id': '5.20', 'tid': None, 'properties': None\}, \{'id': '6.24', 'tid': None, 'properties': None\}])] \ding{56}

			\vspace{0.05cm}
			
			4. \textbf{Query 2:} \textbf{\textcolor[RGB]{0, 128, 0}{MATCH}} (n:nt3 \{p5: 'Ce'\})-[:et3]->(m) \textbf{\textcolor[RGB]{0, 128, 0}{RETURN}} n, m;
			
			\vspace{0.05cm}
			
			5. \textbf{Neo4j:} [Node('nt3', name='u4', p5='Ce'), Node('nt1', name = 'u85', p3=True)], [Node('nt3', name='u4', p5='Ce'), Node('nt2', name='u84', p9=44.79)] \ding{52}
			\vspace{0.05cm}
			
			6. \textbf{Agensgraph:} ('nt3[3.2]\{"p5": "Ce", "name": "u4"\}', 'nt2[5.20]\{"p9": 44.79, "name": "u84"\}'), ('nt3[3.2]\{"p5": "Ce", "name": "u4"\}', 'nt1[6.24]\{"p3": true, "name": "u85"\}') \ding{52}
			\vspace{0.05cm}
			
		}%
	}
	\vspace{0.25cm}
	
	As shown in listing 9, Query 1 first retrieves nodes with the label "nt3" and the property "p5" value 'Ce', then obtains nodes reached along the "et3" relationship from these nodes, and finally aggregates them into a list. Neo4j returns the correct results, while Agensgraph fails to retrieve the node's key-value pairs. Query 2, after removing the "collect" keyword, yields consistent results in both neo4j and Agensgraph. This bug is similar to the one in listing 8, so we suspect that Agensgraph may have a bug when using the "collect" keyword to gather nodes into a list.
	
	\vspace{0.25cm}
	
	\hspace{-0.35cm}\textbf{Listing 10: Cypher queries that trigger a wrong-result bug in Agensgraph 2.13.0.}
	\vspace{0.25cm}

	\hspace{-0.39cm}
	\fbox{%
		\parbox{8.3cm}{
			\vspace{0.05cm}
			1. \textbf{Query 1:} \textbf{\textcolor[RGB]{0, 128, 0}{MATCH}} (n) \textbf{\textcolor[RGB]{0, 128, 0}{WHERE}} n.p2 > 50 \textbf{\textcolor[RGB]{0, 128, 0}{RETURN}} n.name;
			
			\vspace{0.05cm}
			
			2. \textbf{Neo4j:} [] \ding{52}
			\vspace{0.05cm}
			
			3. \textbf{Agensgraph:} [('u19',), ('u56',), ('u96',), ('u33',), ('u44',), ('u47',), ('u86',), ('u17',), ('u37',), ('u91',)] \ding{56}
			\vspace{0.05cm}
			
			4. \textbf{Query 2:} \textbf{\textcolor[RGB]{0, 128, 0}{MATCH}} (n) \textbf{\textcolor[RGB]{0, 128, 0}{RETURN}} n.name, n.p2;
			
			\vspace{0.05cm}
			
			5. \textbf{Neo4j:} [['u17', False], ['u19', True],..., ['u99', None]] \ding{52}
			\vspace{0.05cm}
			
			6. \textbf{Agensgraph:} [('u17', False), ('u19', True),..., ('u99', None)]\ding{52}
			\vspace{0.05cm}
			
		}%
	}
	
	\vspace{0.25cm}
	
	As shown in listing 10, query 1 returns the names of all nodes with the property "p2" greater than 50. Neo4j returns an empty set, while Agensgraph returns the names of some nodes. Query 2 returns the names and "p2" property values of all nodes. We can see that the type of the "p2" property value is a boolean constant, and it cannot be compared with an integer constant. However, Agensgraph still returns results. Therefore, we suspect that Agensgraph may have a bug in setting the priority of boolean constant higher than integer constants when comparing values of different types.

	As shown in listing 11, query 1 aims to return the count of all nodes going out through the "et0" relationship and returning through the "et3" relationship. Neo4j returns 28, while Agensgraph strangely returns a very large value, 2695, because the total number of nodes in the graph is only 100. Query 2 returns the count of nodes going out through the "et0" relationship, and query 3 returns the count of nodes returning through the "et3" relationship. The results of these two queries on both graph database engines are consistent, but we found that the product of the results of queries 2 and 3 equals the result of query 1. From this, we can infer that Agensgraph, when using the above pattern to match multiple relationships, does not consider these relationships simultaneously but matches them separately. Finally, it returns the results in a manner similar to a Cartesian product\cite{44}, leading to a bug.
	
	\vspace{0.25cm}
	
	\hspace{-0.35cm}\textbf{Listing 11: Cypher queries that trigger a wrong-result bug in Agensgraph 2.13.0.}
	\vspace{0.25cm}

	\hspace{-0.39cm}
	\fbox{%
		\parbox{8.3cm}{
			\vspace{0.05cm}
			1. \textbf{Query 1:} \textbf{\textcolor[RGB]{0, 128, 0}{MATCH}} (n)-[:et0]->(), ()-[:et3]->(n) \textbf{\textcolor[RGB]{0, 128, 0}{RETURN count}} (n);
			
			\vspace{0.05cm}
			
			2. \textbf{Neo4j:} [28] \ding{52}    \hspace{9em}             \textbf{Agensgraph:} (2695,) \ding{56}
			\vspace{0.05cm}
			
			3. \textbf{Query 2:} \textbf{\textcolor[RGB]{0, 128, 0}{MATCH}} (n)-[:et0]->() \textbf{\textcolor[RGB]{0, 128, 0}{RETURN count}} (n);
			
			\vspace{0.05cm}
			
			4. \textbf{Neo4j:} [55] \ding{52}    \hspace{9em}             \textbf{Agensgraph:} (55,) \hspace{1em} \ding{52}
			\vspace{0.05cm}

			5. \textbf{Query 3:} \textbf{\textcolor[RGB]{0, 128, 0}{MATCH}} ()-[:et3]->(n) \textbf{\textcolor[RGB]{0, 128, 0}{RETURN count}} (n);
			
			\vspace{0.05cm}
			
			6. \textbf{Neo4j:} [49] \ding{52}     \hspace{9em}         \textbf{Agensgraph:} (49,) \hspace{1em} \ding{52}
			\vspace{0.05cm}

		}%
	}

	\vspace{0.25cm}
	
	\hspace{-0.35cm}\textbf{Listing 12: Cypher queries that trigger a wrong-result bug in Agensgraph 2.13.0.}
	
	\vspace{0.25cm}
	
	\hspace{-0.39cm}
	\fbox{%
		\parbox{8.3cm}{
			\vspace{0.05cm}
			1. \textbf{Query 1:} \textbf{\textcolor[RGB]{0, 128, 0}{MATCH}} (n1)-[r]->(n2:nt0)  \textbf{\textcolor[RGB]{0, 128, 0}{WHERE}} n1.name = 'u9' \textbf{\textcolor[RGB]{0, 128, 0}{RETURN}} \textbf{\textcolor[RGB]{0, 128, 0}{COLLECT}}(\textbf{\textcolor[RGB]{0, 128, 0}{DISTINCT}} n2.p2) \textbf{\textcolor[RGB]{0, 128, 0}{AS}} distinct\_values;
			
			\vspace{0.05cm}
			
			2. \textbf{Neo4j:} [[False]] \ding{52}  \hspace{4em}  
			\textbf{Agensgraph:} ([False, None],) \ding{56}
			\vspace{0.05cm}
			
		}%
	}
	
	\vspace{0.25cm}

	As shown in listing12, this query first matches nodes connected to the 'u9' node with the type "nt0" and then returns the unique set of p2 property values for these nodes. The result returned by Neo4J is 'False', while Agensgraph returns 'False' and 'None'. This is because in Agensgraph, None indicates the absence, meaning that some nodes in the matched nodes may lack the p2 property, resulting in the presence of none. However, Distinct is meant to return unique elements, and after using distinct, None values should be removed to avoid potential user confusion.
	
	\vspace{0.25cm}
	
	\hspace{-0.35cm}\textbf{Listing 13: Gremlin queries that trigger a wrong-result bug in Janusgraph 1.0.0.}
	
	\vspace{0.25cm}

	\hspace{-0.39cm}
	\fbox{%
		\parbox{8.3cm}{
			\vspace{0.05cm}
			1. \textbf{Query 1}: g.\textcolor{orange}{E}().\textcolor{orange}{has}('p2', \textcolor{orange}{without}('GhR')).\textcolor{orange}{count}(); 
			\vspace{0.05cm}
			
			2. \textbf{Janusgraph}:[9] \ding{56}   \hspace{8em}  \textbf{Tinkergraph}:[23] \ding{52}
			
			\vspace{0.05cm}

			3. \textbf{Query 2}: g.\textcolor{orange}{E}().\textcolor{orange}{has}('p2').\textcolor{orange}{count}();
			
			\vspace{0.05cm}
			
			4. \textbf{Janusgraph}:[23] \ding{52}   \hspace{7.5em}  \textbf{Tinkergraph}:[23] \ding{52}

			\vspace{0.05cm}
			
			5. \textbf{Query 3}: g.\textcolor{orange}{E}().\textcolor{orange}{has}('p2', \textcolor{orange}{without}('GhR')).\textcolor{orange}{valueMap}(); 
			\vspace{0.05cm}
			6. \textbf{Janusgraph}: [\{'name': 'e16', 'p2': True\}, \{'p2': True, 'name': 'e13', 'p9': 55.07\},...,\{'name': 'e180', 'p2': True\}] \ding{56}
			
			\vspace{0.05cm}
			7. \textbf{Tinkergraph}:
			[\{'p2': False, 'name': 'e5'\}, \{'p1': '5k', 'p2': False, 'p4': False, 'name': 'e9'\}, \{'p2': True, 'name': 'e16'\},...,\{'p2': True, 'name': 'e13', 'p9': 55.07\}] \ding{52}
			
			\vspace{0.05cm}
		}%
	}
	\vspace{0.25cm}
	
	As shown in listing 13, query 1 returns the number of edges that have the "p2" property and the "p2" property value does not contain 'GhR'. It can be seen that Janusgraph and tinkergraph return different values. Query 2 returns the number of edges that have the "p2" property, and it is observed that the results of the two graph database engines are consistent. Therefore, we believe that there is an issue with the use of the without operator. Subsequently, we use query 3 to return the property key-value pairs of edges that have the "p2" attribute, and the "p2" property value does not contain 'GhR'. We found that Janusgraph only returns edges with "p2" attribute value as True. It can be inferred that Janusgraph has a bug when using the without operator to handle property values of different types.
	
	\vspace{0.25cm}
	
	\hspace{-0.35cm}\textbf{Listing 14: Gremlin queries that trigger a wrong-result bug in Janusgraph 1.0.0.}
	\vspace{0.4cm}
	
	\hspace{-0.39cm}
	\fbox{%
		\parbox{8.3cm}{

			\vspace{0.05cm}
			
			1. result\_set = client1.submit("g.\textcolor{orange}{addV}('nt5').\textcolor{orange}{property}('p9', 13.85 ).\textcolor{orange}{property}('test3', 'pvifo')")
			
			\vspace{0.05cm}
			
			2. nodes = list(map(lambda v: {"label": g.V(v).label().toList(), "properties": g.V(v).valueMap().toList()}, g.V()))
			
			\vspace{0.05cm}
			
			3. for node in nodes:
			
			\vspace{0.05cm}
			
			4. \hspace{0.1em} print("Node Properties:", node["properties"])
			
			\vspace{0.05cm}
			
			5. \textbf{Janusgraph\_Output}: Node Properties: [\{'p9': [13.85], 'test3': ['pvifo']\}]\ding{52}
			
			\vspace{0.05cm}
			
			6. \hspace{-0.02cm}\textbf{Tinkergraph\_Output}: Node Properties: [\{'test3': ['pvifo'], 'p9': [\{'@type': 'gx:BigDecimal', '@value': 13.85\}]\}]\ding{52}
			
			\vspace{0.05cm}
			
			7. g.V().drop().iterate(); g.E().drop().iterate();
			
			\vspace{0.05cm}
			
			8. result\_set = client1.submit("g.\textcolor{orange}{addV}('nt5').\textcolor{orange}{property}('p9', 13 ).\textcolor{orange}{property}('test3', 25.6)")
			
			\vspace{0.05cm}
			
			9. nodes = list(map(lambda v: {"label": g.V(v).label().toList(), "properties": g.V(v).valueMap().toList()}, g.V()))
			
			\vspace{0.05cm}
			
			10. for node in nodes:
			
			\vspace{0.05cm}
			
			11. \hspace{0.2em}print("Node Properties:", node["properties"])
			
			\vspace{0.05cm}
			
			12. \textbf{Janusgraph\_Output}: Node Properties: [\{'p9': [13.0], 'test3': ['25.6']\}]  \ding{56}
			
			\vspace{0.10cm}

			13. \textbf{Tinkergraph\_Output}: Node Properties: [\{'test3': [\{'@type': 'gx:BigDecimal', '@value': 25.6\}], 'p9': [13]\}]  \ding{52}
			
			\vspace{0.05cm}
			
		}%
	}
	
	\vspace{0.4cm}
	
	As shown in listing 14, we provide a partial code example in Python to create graph data. Lines 1-4 create a node and return its property key-value pairs. Line 7 deletes all nodes and edges from the graph. Lines 8-11 create a node with the same type but different property types and return its property key-value pairs. From the output results in lines 5 and 6, it can be seen that the property values of nodes constructed under different graph database engines are the same. From the result in line 12, it can be seen that Janusgraph creates the 'p9' property value and 'test3' property value of the node with integer and float, but displays them as float and string. Line 13 shows that Tinkergraph displays property values in a type consistent with the input during node creation. Therefore, we infer that even after deleting all nodes and edges from the graph, the property value types of the nodes initially constructed in Janusgraph will still affect the types of subsequently constructed node property value, leading to a bug.

	As shown in listing 15, query 1 returns the key-value pairs of nodes with the 'p2' property and 'p2' property value not equal to the string 'false'.  The node property key-value pairs returned by Janusgraph are far fewer than those returned by Tinkergraph. Query 2 returns the key-value pairs of nodes with the 'p2' property and 'p2' property value not equal to the boolean value false, and it can be seen that the results returned by Janusgraph and Tinkergraph are consistent. By comparing the results of query 1 and query 2 in Janusgraph, we can conclude that Janusgraph treats the string 'false' property value as a boolean value false, leading to this bug.

	\vspace{0.25cm}

	\hspace{-0.35cm}\textbf{Listing 15: Gremlin queries that trigger a wrong-result bug in Janusgraph 1.0.0.}
	\vspace{0.25cm}

	\hspace{-0.39cm}
	\fbox{%
		\parbox{8.3cm}{
			
			\vspace{0.05cm}
			
			1. \textbf{Query 1}: g.\textcolor{orange}{V}().\textcolor{orange}{has}('p2', \textcolor{orange}{neq}('false')).\textcolor{orange}{valueMap}(); 
			
			\vspace{0.05cm}
			
			2. \textbf{Janusgraph}:[{'p2': [True], 'name': ['u96']}, {'p2': [True], 'name': ['u19']}, {'p2': [True], 'name': ['u47']}] \ding{56}
			
			\vspace{0.05cm}
			
			3. \textbf{Tinkergraph}:[{'p2': [False], 'name': ['u91']}, {'p2': [True], 'name': ['u96']},...,{'p2': [True], 'name': ['u47']}] \ding{52}
			
			\vspace{0.05cm}

			4. \textbf{Query 2}: g.\textcolor{orange}{V}().\textcolor{orange}{has}('p2', \textcolor{orange}{neq}(false)).\textcolor{orange}{valueMap}(); 
			
			\vspace{0.05cm}
			
			5. \textbf{Janusgraph}:[{'p2': [True], 'name': ['u96']}, {'p2': [True], 'name': ['u19']}, {'p2': [True], 'name': ['u47']}] \ding{52}
			
			\vspace{0.05cm}
			
			6. \textbf{Tinkergraph}:[{'p2': [True], 'name': ['u96']}, {'p2': [True], 'name': ['u19']}, {'p2': [True], 'name': ['u47']}] \ding{52}
			
			\vspace{0.05cm}

		}%
	}
	
	\vspace{0.25cm}

	From the above bug analysis, it can be seen that bugs in graph database engines mainly occur in the comparison of different types of property values (e.g., Listing 10, Listing 13), the use of certain operators (e.g., Listing 8, Listing 9), execution order under complex queries (e.g., Listing 6), default settings for null and boolean values (e.g., Listing 7, Listing 12, Listing 15), handling logic under complex pattern matching (e.g., Listing 11), and incomplete execution of some commands (e.g., Listing 14). The above-mentioned bugs have all been submitted to the corresponding communities of the graph database engines on github.

	\section{RELATED WORK}
	Recently, researchers have proposed various methods for detecting bugs in relational database engines. SQLsmith\cite{21} is a fuzz testing method for detecting bugs in relational database engines. It first generates SQL query randomly according to the Abstract Syntax Tree (AST)  generator, then simplifies the generated SQL query through SQL Reducer. Finally, it determines the presence of bugs based on whether the generated SQL query can successfully run on the relational database engines. While this method is effective in detecting crash bugs, it cannot identify wrong-result bugs. PQS\cite{46} selects a target data from randomly generated tables, generates conditional expressions based on the target data, constructs an SQL query with a where or join clause, and determines the presence of bugs by checking if the result is included in the result set.  While this method can effectively detect bugs in relational database engines, its applicability is limited across different relational database engines due to variations in SQL syntax and query optimization strategies. NoRec\cite{47} converts the original SQL query into a non-optimized SQL query and compares the results of these two SQL queries for consistency.  However, this method cannot detect bugs in SQL queries with subquery structures. TLP\cite{37} transforms randomly generated original SQL queries into three different logical queries based on the true, false, and null ternary logic. It detects bugs in relational database engines by comparing the result sets of the transformed three SQL queries with the original SQL query. TQS\cite{48} proposes a testing framework capable of detecting logical bugs arising from multi-table join queries. The authors treat the database schema as a graph and use biased random walks to generate SQL queries. Based on biased random walks and graph embedding methods, high-quality SQL queries are generated to detect numerous logical bugs in relational database engines.
	
	In contrast to detecting bugs in relational database engines, the identification of bugs in graph database engines is relatively less discussed. This is due to the significantly earlier development of relational database engines compared to graph database engines. However, significant differences exist in data models, query languages, storage structures, etc., between relational and graph database engines. As a result, bug detection methods designed for relational database engines cannot be directly applied to graph database engines. Moreover, because graph database engines use nodes and edges to represent entities and their relationships, the data model of graph database engines becomes more flexible and complex. For instance, nodes can have an arbitrary number of properties and relationships, increasing the complexity of bug detection in graph database engines. Furthermore, unlike relational database engines that have a universal and standardized query language like SQL\cite{49}, each graph database engine has its proprietary query language. For instance, AgensGraph uses Cypher, and JanusGraph uses Gremlin, making it challenging to design a bug detection method that is universally applicable across different graph database engines.

	The detection method for bugs in graph database engines primarily employs differential testing\cite{50}. By running identical queries on different graph database engines or different versions of graph database engines and comparing their query results, this type of method can effectively identify bugs in graph database engines based on result inconsistencies. For instance, Grand\cite{18} constructed a traversal model to generate syntactically correct and meaningful Gremlin queries. These queries were then input into various graph database engines based on the Gremlin query language, and the results of queries across different graph database engines were compared. GDSmith\cite{19}, guided by property graphs, generates Cypher queries, inputs them into  graph database engines based on the Cypher query language, and compares the query results across different graph database engines. In contrast to the aforementioned methods, DGDB, based on the recently popularized large language model ChatGPT, proposes a paradigm for detecting bugs in graph database engines. This paradigm, without requiring extensive expert knowledge, is capable of detecting bugs in graph database engines based on different query languages (including but not limited to Cypher, Gremlin) and can increase the non-empty-result ratio of generated queries, significantly enhancing the efficiency of detecting graph database engine bugs.
	
	\vspace{-0.1cm}
	
	\section{CONCLUSION}
	
	In this paper, we propose a simple paradigm DGDB for detecting graph database engine bugs. This paradigm offers a unique perspective in query generation for bug detection by integrating the cutting-edge Large Language Models with graph query languages. It firstly leverages powerful capabilities of ChatGPT in understanding and generating natural language to generate queries that meet specified constraints. Subsequently, it uses differential testing method to detect graph database engine bugs. We applied the paradigm to cypher-based graph database engines and gremlin-based graph database engines, proposing DGDB-cypher and DGDB-gremlin to detect bugs in graph database engines. Experimental results demonstrate that our method is effective in discovering bugs in the latest versions of graph database engines for both query languages.

	\bibliographystyle{ACM-Reference-Format}
	\bibliography{sample-base}


\begin{thebibliography}{50}


\ifx \showCODEN    \undefined \def \showCODEN     #1{\unskip}     \fi
\ifx \showDOI      \undefined \def \showDOI       #1{#1}\fi
\ifx \showISBNx    \undefined \def \showISBNx     #1{\unskip}     \fi
\ifx \showISBNxiii \undefined \def \showISBNxiii  #1{\unskip}     \fi
\ifx \showISSN     \undefined \def \showISSN      #1{\unskip}     \fi
\ifx \showLCCN     \undefined \def \showLCCN      #1{\unskip}     \fi
\ifx \shownote     \undefined \def \shownote      #1{#1}          \fi
\ifx \showarticletitle \undefined \def \showarticletitle #1{#1}   \fi
\ifx \showURL      \undefined \def \showURL       {\relax}        \fi
\providecommand\bibfield[2]{#2}
\providecommand\bibinfo[2]{#2}
\providecommand\natexlab[1]{#1}
\providecommand\showeprint[2][]{arXiv:#2}

\bibitem[Abdelaziz et~al\mbox{.}(2017)]%
        {28}
\bibfield{author}{\bibinfo{person}{Ibrahim Abdelaziz}, \bibinfo{person}{Essam
  Mansour}, \bibinfo{person}{Mourad Ouzzani}, \bibinfo{person}{Ashraf
  Aboulnaga}, {and} \bibinfo{person}{Panos Kalnis}.}
  \bibinfo{year}{2017}\natexlab{}.
\newblock \showarticletitle{Query optimizations over decentralized RDF graphs}.
  In \bibinfo{booktitle}{\emph{2017 IEEE 33rd International Conference on Data
  Engineering (ICDE)}}. IEEE, \bibinfo{pages}{139--142}.
\newblock


\bibitem[Amazon(2023)]%
        {7}
\bibfield{author}{\bibinfo{person}{DataStax Dylan Bethune-Waddell Expero Google
  Orion Health IBM Rafael Fernandes Robert Dale~Seeq Amazon, Aurelius}.}
  \bibinfo{year}{2023}\natexlab{}.
\newblock \bibinfo{title}{JanusGraph}.
\newblock \bibinfo{howpublished}{\url{https://janusgraph.org/}}.
\newblock


\bibitem[Andreas~Seltenreich(2022)]%
        {21}
\bibfield{author}{\bibinfo{person}{Sjoerd~Mullender Andreas~Seltenreich,
  Bo~Tang}.} \bibinfo{year}{2022}\natexlab{}.
\newblock \bibinfo{title}{SQLsmith}.
\newblock \bibinfo{howpublished}{\url{https://github.com/anse1/sqlsmith}}.
\newblock


\bibitem[Angles and Gutierrez(2008)]%
        {4}
\bibfield{author}{\bibinfo{person}{Renzo Angles} {and} \bibinfo{person}{Claudio
  Gutierrez}.} \bibinfo{year}{2008}\natexlab{}.
\newblock \showarticletitle{Survey of graph database models}.
\newblock \bibinfo{journal}{\emph{ACM Computing Surveys (CSUR)}}
  \bibinfo{volume}{40}, \bibinfo{number}{1} (\bibinfo{year}{2008}),
  \bibinfo{pages}{1--39}.
\newblock


\bibitem[Angluin(1990)]%
        {36}
\bibfield{author}{\bibinfo{person}{Dana Angluin}.}
  \bibinfo{year}{1990}\natexlab{}.
\newblock \showarticletitle{Negative results for equivalence queries}.
\newblock \bibinfo{journal}{\emph{Machine Learning}}  \bibinfo{volume}{5}
  (\bibinfo{year}{1990}), \bibinfo{pages}{121--150}.
\newblock


\bibitem[Brown et~al\mbox{.}(2020)]%
        {29}
\bibfield{author}{\bibinfo{person}{Tom Brown}, \bibinfo{person}{Benjamin Mann},
  \bibinfo{person}{Nick Ryder}, \bibinfo{person}{Melanie Subbiah},
  \bibinfo{person}{Jared~D Kaplan}, \bibinfo{person}{Prafulla Dhariwal},
  \bibinfo{person}{Arvind Neelakantan}, \bibinfo{person}{Pranav Shyam},
  \bibinfo{person}{Girish Sastry}, \bibinfo{person}{Amanda Askell},
  {et~al\mbox{.}}} \bibinfo{year}{2020}\natexlab{}.
\newblock \showarticletitle{Language models are few-shot learners}.
\newblock \bibinfo{journal}{\emph{Advances in neural information processing
  systems}}  \bibinfo{volume}{33} (\bibinfo{year}{2020}),
  \bibinfo{pages}{1877--1901}.
\newblock


\bibitem[Chamberlin and Boyce(1974)]%
        {49}
\bibfield{author}{\bibinfo{person}{Donald~D Chamberlin} {and}
  \bibinfo{person}{Raymond~F Boyce}.} \bibinfo{year}{1974}\natexlab{}.
\newblock \showarticletitle{SEQUEL: A structured English query language}. In
  \bibinfo{booktitle}{\emph{Proceedings of the 1974 ACM SIGFIDET (now SIGMOD)
  workshop on Data description, access and control}}.
  \bibinfo{pages}{249--264}.
\newblock


\bibitem[Chen et~al\mbox{.}(2023)]%
        {26}
\bibfield{author}{\bibinfo{person}{Zihan Chen}, \bibinfo{person}{Lei~Nico
  Zheng}, \bibinfo{person}{Cheng Lu}, \bibinfo{person}{Jialu Yuan}, {and}
  \bibinfo{person}{Di Zhu}.} \bibinfo{year}{2023}\natexlab{}.
\newblock \showarticletitle{ChatGPT Informed Graph Neural Network for Stock
  Movement Prediction}.
\newblock \bibinfo{journal}{\emph{arXiv preprint arXiv:2306.03763}}
  (\bibinfo{year}{2023}).
\newblock


\bibitem[Codd(1983)]%
        {1}
\bibfield{author}{\bibinfo{person}{Edgar~Frank Codd}.}
  \bibinfo{year}{1983}\natexlab{}.
\newblock \showarticletitle{A relational model of data for large shared data
  banks}.
\newblock \bibinfo{journal}{\emph{Commun. ACM}} \bibinfo{volume}{26},
  \bibinfo{number}{1} (\bibinfo{year}{1983}), \bibinfo{pages}{64--69}.
\newblock


\bibitem[Cui et~al\mbox{.}(2023)]%
        {24}
\bibfield{author}{\bibinfo{person}{Jiaxi Cui}, \bibinfo{person}{Zongjian Li},
  \bibinfo{person}{Yang Yan}, \bibinfo{person}{Bohua Chen}, {and}
  \bibinfo{person}{Li Yuan}.} \bibinfo{year}{2023}\natexlab{}.
\newblock \showarticletitle{Chatlaw: Open-source legal large language model
  with integrated external knowledge bases}.
\newblock \bibinfo{journal}{\emph{arXiv preprint arXiv:2306.16092}}
  (\bibinfo{year}{2023}).
\newblock


\bibitem[Czerepicki(2016)]%
        {15}
\bibfield{author}{\bibinfo{person}{Andrzej Czerepicki}.}
  \bibinfo{year}{2016}\natexlab{}.
\newblock \showarticletitle{Application of graph databases for transport
  purposes}.
\newblock \bibinfo{journal}{\emph{Bulletin of the Polish Academy of Sciences.
  Technical Sciences}} \bibinfo{volume}{64}, \bibinfo{number}{3}
  (\bibinfo{year}{2016}), \bibinfo{pages}{457--466}.
\newblock


\bibitem[Enterprise(2023)]%
        {8}
\bibfield{author}{\bibinfo{person}{SAP Enterprise}.}
  \bibinfo{year}{2023}\natexlab{}.
\newblock \bibinfo{title}{OrientDB Community Edition}.
\newblock \bibinfo{howpublished}{\url{http://www.orientdb.org/}}.
\newblock


\bibitem[Evans and Savoia(2007)]%
        {50}
\bibfield{author}{\bibinfo{person}{Robert~B Evans} {and}
  \bibinfo{person}{Alberto Savoia}.} \bibinfo{year}{2007}\natexlab{}.
\newblock \showarticletitle{Differential testing: a new approach to change
  detection}. In \bibinfo{booktitle}{\emph{The 6th Joint Meeting on European
  software engineering conference and the ACM SIGSOFT Symposium on the
  Foundations of Software Engineering: Companion Papers}}.
  \bibinfo{pages}{549--552}.
\newblock


\bibitem[Foundation(2023a)]%
        {9}
\bibfield{author}{\bibinfo{person}{Apache~Software Foundation}.}
  \bibinfo{year}{2023}\natexlab{a}.
\newblock \bibinfo{title}{HugeGraph}.
\newblock \bibinfo{howpublished}{\url{https://hugegraph.apache.org/}}.
\newblock


\bibitem[Foundation(2023b)]%
        {10}
\bibfield{author}{\bibinfo{person}{Apache~Software Foundation}.}
  \bibinfo{year}{2023}\natexlab{b}.
\newblock \bibinfo{title}{TinkerGraph}.
\newblock
  \bibinfo{howpublished}{\url{https://tinkerpop.incubator.apache.org/}}.
\newblock


\bibitem[Francis et~al\mbox{.}(2018)]%
        {33}
\bibfield{author}{\bibinfo{person}{Nadime Francis}, \bibinfo{person}{Alastair
  Green}, \bibinfo{person}{Paolo Guagliardo}, \bibinfo{person}{Leonid Libkin},
  \bibinfo{person}{Tobias Lindaaker}, \bibinfo{person}{Victor Marsault},
  \bibinfo{person}{Stefan Plantikow}, \bibinfo{person}{Mats Rydberg},
  \bibinfo{person}{Petra Selmer}, {and} \bibinfo{person}{Andr{\'e}s Taylor}.}
  \bibinfo{year}{2018}\natexlab{}.
\newblock \showarticletitle{Cypher: An evolving query language for property
  graphs}. In \bibinfo{booktitle}{\emph{Proceedings of the 2018 international
  conference on management of data}}. \bibinfo{pages}{1433--1445}.
\newblock


\bibitem[Gao et~al\mbox{.}(2015)]%
        {45}
\bibfield{author}{\bibinfo{person}{Qing Gao}, \bibinfo{person}{Hansheng Zhang},
  \bibinfo{person}{Jie Wang}, \bibinfo{person}{Yingfei Xiong},
  \bibinfo{person}{Lu Zhang}, {and} \bibinfo{person}{Hong Mei}.}
  \bibinfo{year}{2015}\natexlab{}.
\newblock \showarticletitle{Fixing recurring crash bugs via analyzing q\&a
  sites (T)}. In \bibinfo{booktitle}{\emph{2015 30th IEEE/ACM International
  Conference on Automated Software Engineering (ASE)}}. IEEE,
  \bibinfo{pages}{307--318}.
\newblock


\bibitem[Guillermo et~al\mbox{.}(2022)]%
        {16}
\bibfield{author}{\bibinfo{person}{Marielet Guillermo},
  \bibinfo{person}{Maverick Rivera}, \bibinfo{person}{Ronnie Concepcion},
  \bibinfo{person}{Robert~Kerwin Billones}, \bibinfo{person}{Argel Bandala},
  \bibinfo{person}{Edwin Sybingco}, \bibinfo{person}{Alexis Fillone}, {and}
  \bibinfo{person}{Elmer Dadios}.} \bibinfo{year}{2022}\natexlab{}.
\newblock \showarticletitle{Graph Database-modelled Public Transportation Data
  for Geographic Insight Web Application}. In \bibinfo{booktitle}{\emph{2022
  IEEE/ACIS 23rd International Conference on Software Engineering, Artificial
  Intelligence, Networking and Parallel/Distributed Computing (SNPD)}}. IEEE,
  \bibinfo{pages}{2--7}.
\newblock


\bibitem[Hewitt and Savage(1955)]%
        {44}
\bibfield{author}{\bibinfo{person}{Edwin Hewitt} {and}
  \bibinfo{person}{Leonard~J Savage}.} \bibinfo{year}{1955}\natexlab{}.
\newblock \showarticletitle{Symmetric measures on Cartesian products}.
\newblock \bibinfo{journal}{\emph{Trans. Amer. Math. Soc.}}
  \bibinfo{volume}{80}, \bibinfo{number}{2} (\bibinfo{year}{1955}),
  \bibinfo{pages}{470--501}.
\newblock


\bibitem[Hua et~al\mbox{.}(2023)]%
        {19}
\bibfield{author}{\bibinfo{person}{Ziyue Hua}, \bibinfo{person}{Wei Lin},
  \bibinfo{person}{Luyao Ren}, \bibinfo{person}{Zongyang Li},
  \bibinfo{person}{Lu Zhang}, \bibinfo{person}{Wenpin Jiao}, {and}
  \bibinfo{person}{Tao Xie}.} \bibinfo{year}{2023}\natexlab{}.
\newblock \showarticletitle{GDsmith: Detecting bugs in Cypher graph database
  engines}. In \bibinfo{booktitle}{\emph{Proceedings of ACM SIGSOFT
  International Symposium on Software Testing and Analysis (ISSTA)}}.
\newblock


\bibitem[Inc(2023a)]%
        {11}
\bibfield{author}{\bibinfo{person}{Bitnine~Global Inc}.}
  \bibinfo{year}{2023}\natexlab{a}.
\newblock \bibinfo{title}{Agensgraph}.
\newblock \bibinfo{howpublished}{\url{https://bitnine.net/agensgraph/}}.
\newblock


\bibitem[Inc(2023b)]%
        {6}
\bibfield{author}{\bibinfo{person}{Neo4j Inc}.}
  \bibinfo{year}{2023}\natexlab{b}.
\newblock \bibinfo{title}{Neo4j}.
\newblock \bibinfo{howpublished}{\url{https://neo4j.com/}}.
\newblock


\bibitem[Jain et~al\mbox{.}(2005)]%
        {35}
\bibfield{author}{\bibinfo{person}{Himanshu Jain}, \bibinfo{person}{Daniel
  Kroening}, \bibinfo{person}{Natasha Sharygina}, {and} \bibinfo{person}{Edmund
  Clarke}.} \bibinfo{year}{2005}\natexlab{}.
\newblock \showarticletitle{Word level predicate abstraction and refinement for
  verifying RTL Verilog}. In \bibinfo{booktitle}{\emph{Proceedings of the 42nd
  annual Design Automation Conference}}. \bibinfo{pages}{445--450}.
\newblock


\bibitem[Kamm(2022)]%
        {17}
\bibfield{author}{\bibinfo{person}{Matteo Kamm}.}
  \bibinfo{year}{2022}\natexlab{}.
\newblock \emph{\bibinfo{title}{Testing Graph Databases using Predicate
  Partitioning}}.
\newblock \bibinfo{thesistype}{Master's\ thesis}. \bibinfo{school}{ETH Zurich}.
\newblock


\bibitem[Kirillov et~al\mbox{.}(2023)]%
        {23}
\bibfield{author}{\bibinfo{person}{Alexander Kirillov}, \bibinfo{person}{Eric
  Mintun}, \bibinfo{person}{Nikhila Ravi}, \bibinfo{person}{Hanzi Mao},
  \bibinfo{person}{Chloe Rolland}, \bibinfo{person}{Laura Gustafson},
  \bibinfo{person}{Tete Xiao}, \bibinfo{person}{Spencer Whitehead},
  \bibinfo{person}{Alexander~C Berg}, \bibinfo{person}{Wan-Yen Lo},
  {et~al\mbox{.}}} \bibinfo{year}{2023}\natexlab{}.
\newblock \showarticletitle{Segment anything}.
\newblock \bibinfo{journal}{\emph{arXiv preprint arXiv:2304.02643}}
  (\bibinfo{year}{2023}).
\newblock


\bibitem[Konno et~al\mbox{.}(2017)]%
        {12}
\bibfield{author}{\bibinfo{person}{Takahiro Konno}, \bibinfo{person}{Runhe
  Huang}, \bibinfo{person}{Tao Ban}, {and} \bibinfo{person}{Chuanhe Huang}.}
  \bibinfo{year}{2017}\natexlab{}.
\newblock \showarticletitle{Goods recommendation based on retail knowledge in a
  Neo4j graph database combined with an inference mechanism implemented in
  jess}. In \bibinfo{booktitle}{\emph{2017 IEEE SmartWorld, Ubiquitous
  Intelligence \& Computing, Advanced \& Trusted Computed, Scalable Computing
  \& Communications, Cloud \& Big Data Computing, Internet of People and Smart
  City Innovation (SmartWorld/SCALCOM/UIC/ATC/CBDCom/IOP/SCI)}}. IEEE,
  \bibinfo{pages}{1--8}.
\newblock


\bibitem[Luo et~al\mbox{.}(2018)]%
        {42}
\bibfield{author}{\bibinfo{person}{Bingfeng Luo}, \bibinfo{person}{Yansong
  Feng}, \bibinfo{person}{Zheng Wang}, \bibinfo{person}{Songfang Huang},
  \bibinfo{person}{Rui Yan}, {and} \bibinfo{person}{Dongyan Zhao}.}
  \bibinfo{year}{2018}\natexlab{}.
\newblock \showarticletitle{Marrying up regular expressions with neural
  networks: A case study for spoken language understanding}.
\newblock \bibinfo{journal}{\emph{arXiv preprint arXiv:1805.05588}}
  (\bibinfo{year}{2018}).
\newblock


\bibitem[Lyu et~al\mbox{.}(2023)]%
        {25}
\bibfield{author}{\bibinfo{person}{Hanjia Lyu}, \bibinfo{person}{Song Jiang},
  \bibinfo{person}{Hanqing Zeng}, \bibinfo{person}{Yinglong Xia}, {and}
  \bibinfo{person}{Jiebo Luo}.} \bibinfo{year}{2023}\natexlab{}.
\newblock \showarticletitle{Llm-rec: Personalized recommendation via prompting
  large language models}.
\newblock \bibinfo{journal}{\emph{arXiv preprint arXiv:2307.15780}}
  (\bibinfo{year}{2023}).
\newblock


\bibitem[Ma and Wang(2023)]%
        {41}
\bibfield{author}{\bibinfo{person}{Jun Ma} {and} \bibinfo{person}{Bo Wang}.}
  \bibinfo{year}{2023}\natexlab{}.
\newblock \showarticletitle{Segment anything in medical images}.
\newblock \bibinfo{journal}{\emph{arXiv preprint arXiv:2304.12306}}
  (\bibinfo{year}{2023}).
\newblock


\bibitem[McKeeman(1998)]%
        {32}
\bibfield{author}{\bibinfo{person}{William~M McKeeman}.}
  \bibinfo{year}{1998}\natexlab{}.
\newblock \showarticletitle{Differential testing for software}.
\newblock \bibinfo{journal}{\emph{Digital Technical Journal}}
  \bibinfo{volume}{10}, \bibinfo{number}{1} (\bibinfo{year}{1998}),
  \bibinfo{pages}{100--107}.
\newblock


\bibitem[OpenAI(2022)]%
        {38}
\bibfield{author}{\bibinfo{person}{OpenAI}.} \bibinfo{year}{2022}\natexlab{}.
\newblock \bibinfo{title}{ChatGPT}.
\newblock \bibinfo{howpublished}{\url{https://openai.com/blog/chatgpt/}}.
\newblock


\bibitem[Ouyang et~al\mbox{.}(2022)]%
        {30}
\bibfield{author}{\bibinfo{person}{Long Ouyang}, \bibinfo{person}{Jeffrey Wu},
  \bibinfo{person}{Xu Jiang}, \bibinfo{person}{Diogo Almeida},
  \bibinfo{person}{Carroll Wainwright}, \bibinfo{person}{Pamela Mishkin},
  \bibinfo{person}{Chong Zhang}, \bibinfo{person}{Sandhini Agarwal},
  \bibinfo{person}{Katarina Slama}, \bibinfo{person}{Alex Ray},
  {et~al\mbox{.}}} \bibinfo{year}{2022}\natexlab{}.
\newblock \showarticletitle{Training language models to follow instructions
  with human feedback}.
\newblock \bibinfo{journal}{\emph{Advances in Neural Information Processing
  Systems}}  \bibinfo{volume}{35} (\bibinfo{year}{2022}),
  \bibinfo{pages}{27730--27744}.
\newblock


\bibitem[Reagan and Reagan(2018)]%
        {3}
\bibfield{author}{\bibinfo{person}{Rob Reagan} {and} \bibinfo{person}{Rob
  Reagan}.} \bibinfo{year}{2018}\natexlab{}.
\newblock \showarticletitle{Cosmos DB}.
\newblock \bibinfo{journal}{\emph{Web Applications on Azure: Developing for
  Global Scale}} (\bibinfo{year}{2018}), \bibinfo{pages}{187--255}.
\newblock


\bibitem[Ren et~al\mbox{.}(2021)]%
        {14}
\bibfield{author}{\bibinfo{person}{Yuxiang Ren}, \bibinfo{person}{Hao Zhu},
  \bibinfo{person}{Jiawei Zhang}, \bibinfo{person}{Peng Dai}, {and}
  \bibinfo{person}{Liefeng Bo}.} \bibinfo{year}{2021}\natexlab{}.
\newblock \showarticletitle{Ensemfdet: An ensemble approach to fraud detection
  based on bipartite graph}. In \bibinfo{booktitle}{\emph{2021 IEEE 37th
  International Conference on Data Engineering (ICDE)}}. IEEE,
  \bibinfo{pages}{2039--2044}.
\newblock


\bibitem[Rigger(2022)]%
        {20}
\bibfield{author}{\bibinfo{person}{Manuel Rigger}.}
  \bibinfo{year}{2022}\natexlab{}.
\newblock \bibinfo{title}{SQLancer}.
\newblock \bibinfo{howpublished}{\url{https://github.com/sqlancer/sqlancer}}.
\newblock


\bibitem[Rigger and Su(2020a)]%
        {47}
\bibfield{author}{\bibinfo{person}{Manuel Rigger} {and}
  \bibinfo{person}{Zhendong Su}.} \bibinfo{year}{2020}\natexlab{a}.
\newblock \showarticletitle{Detecting optimization bugs in database engines via
  non-optimizing reference engine construction}. In
  \bibinfo{booktitle}{\emph{Proceedings of the 28th ACM Joint Meeting on
  European Software Engineering Conference and Symposium on the Foundations of
  Software Engineering}}. \bibinfo{pages}{1140--1152}.
\newblock


\bibitem[Rigger and Su(2020b)]%
        {37}
\bibfield{author}{\bibinfo{person}{Manuel Rigger} {and}
  \bibinfo{person}{Zhendong Su}.} \bibinfo{year}{2020}\natexlab{b}.
\newblock \showarticletitle{Finding bugs in database systems via query
  partitioning}.
\newblock \bibinfo{journal}{\emph{Proceedings of the ACM on Programming
  Languages}} \bibinfo{volume}{4}, \bibinfo{number}{OOPSLA}
  (\bibinfo{year}{2020}), \bibinfo{pages}{1--30}.
\newblock


\bibitem[Rigger and Su(2020c)]%
        {46}
\bibfield{author}{\bibinfo{person}{Manuel Rigger} {and}
  \bibinfo{person}{Zhendong Su}.} \bibinfo{year}{2020}\natexlab{c}.
\newblock \showarticletitle{Testing database engines via pivoted query
  synthesis}. In \bibinfo{booktitle}{\emph{14th USENIX Symposium on Operating
  Systems Design and Implementation (OSDI 20)}}. \bibinfo{pages}{667--682}.
\newblock


\bibitem[Robinson et~al\mbox{.}(2015)]%
        {2}
\bibfield{author}{\bibinfo{person}{Ian Robinson}, \bibinfo{person}{Jim Webber},
  {and} \bibinfo{person}{Emil Eifrem}.} \bibinfo{year}{2015}\natexlab{}.
\newblock \bibinfo{booktitle}{\emph{Graph databases: new opportunities for
  connected data}}.
\newblock \bibinfo{publisher}{" O'Reilly Media, Inc."}.
\newblock


\bibitem[Rodriguez(2015a)]%
        {34}
\bibfield{author}{\bibinfo{person}{Marko~A Rodriguez}.}
  \bibinfo{year}{2015}\natexlab{a}.
\newblock \showarticletitle{The gremlin graph traversal machine and language
  (invited talk)}. In \bibinfo{booktitle}{\emph{Proceedings of the 15th
  Symposium on Database Programming Languages}}. \bibinfo{pages}{1--10}.
\newblock


\bibitem[Rodriguez(2015b)]%
        {43}
\bibfield{author}{\bibinfo{person}{Marko~A Rodriguez}.}
  \bibinfo{year}{2015}\natexlab{b}.
\newblock \showarticletitle{The gremlin graph traversal machine and language
  (invited talk)}. In \bibinfo{booktitle}{\emph{Proceedings of the 15th
  Symposium on Database Programming Languages}}. \bibinfo{pages}{1--10}.
\newblock


\bibitem[Sen et~al\mbox{.}(2021)]%
        {13}
\bibfield{author}{\bibinfo{person}{Sudipta Sen}, \bibinfo{person}{Akash Mehta},
  \bibinfo{person}{Runa Ganguli}, {and} \bibinfo{person}{Soumya Sen}.}
  \bibinfo{year}{2021}\natexlab{}.
\newblock \showarticletitle{Recommendation of influenced products using
  association rule mining: Neo4j as a case study}.
\newblock \bibinfo{journal}{\emph{SN Computer Science}}  \bibinfo{volume}{2}
  (\bibinfo{year}{2021}), \bibinfo{pages}{1--17}.
\newblock


\bibitem[Slutz(1998)]%
        {22}
\bibfield{author}{\bibinfo{person}{Donald~R Slutz}.}
  \bibinfo{year}{1998}\natexlab{}.
\newblock \showarticletitle{Massive stochastic testing of SQL}. In
  \bibinfo{booktitle}{\emph{VLDB}}, Vol.~\bibinfo{volume}{98}. Citeseer,
  \bibinfo{pages}{618--622}.
\newblock


\bibitem[solid IT(2023)]%
        {5}
\bibfield{author}{\bibinfo{person}{solid IT}.} \bibinfo{year}{2023}\natexlab{}.
\newblock \bibinfo{title}{DB-Engines Ranking of Graph DBMS}.
\newblock
  \bibinfo{howpublished}{\url{https://db-engines.com/en/ranking/graph+dbms}}.
\newblock


\bibitem[Tang et~al\mbox{.}(2023)]%
        {48}
\bibfield{author}{\bibinfo{person}{Xiu Tang}, \bibinfo{person}{Sai Wu},
  \bibinfo{person}{Dongxiang Zhang}, \bibinfo{person}{Feifei Li}, {and}
  \bibinfo{person}{Gang Chen}.} \bibinfo{year}{2023}\natexlab{}.
\newblock \showarticletitle{Detecting Logic Bugs of Join Optimizations in
  DBMS}.
\newblock \bibinfo{journal}{\emph{Proceedings of the ACM on Management of
  Data}} \bibinfo{volume}{1}, \bibinfo{number}{1} (\bibinfo{year}{2023}),
  \bibinfo{pages}{1--26}.
\newblock


\bibitem[Wang et~al\mbox{.}(2020)]%
        {27}
\bibfield{author}{\bibinfo{person}{Ran Wang}, \bibinfo{person}{Zhengyi Yang},
  \bibinfo{person}{Wenjie Zhang}, {and} \bibinfo{person}{Xuemin Lin}.}
  \bibinfo{year}{2020}\natexlab{}.
\newblock \showarticletitle{An empirical study on recent graph database
  systems}. In \bibinfo{booktitle}{\emph{Knowledge Science, Engineering and
  Management: 13th International Conference, KSEM 2020, Hangzhou, China, August
  28--30, 2020, Proceedings, Part I 13}}. Springer, \bibinfo{pages}{328--340}.
\newblock


\bibitem[Wu et~al\mbox{.}(2020)]%
        {40}
\bibfield{author}{\bibinfo{person}{Zonghan Wu}, \bibinfo{person}{Shirui Pan},
  \bibinfo{person}{Fengwen Chen}, \bibinfo{person}{Guodong Long},
  \bibinfo{person}{Chengqi Zhang}, {and} \bibinfo{person}{S~Yu Philip}.}
  \bibinfo{year}{2020}\natexlab{}.
\newblock \showarticletitle{A comprehensive survey on graph neural networks}.
\newblock \bibinfo{journal}{\emph{IEEE transactions on neural networks and
  learning systems}} \bibinfo{volume}{32}, \bibinfo{number}{1}
  (\bibinfo{year}{2020}), \bibinfo{pages}{4--24}.
\newblock


\bibitem[Zhang(2023)]%
        {31}
\bibfield{author}{\bibinfo{person}{Jiawei Zhang}.}
  \bibinfo{year}{2023}\natexlab{}.
\newblock \showarticletitle{Graph-ToolFormer: To Empower LLMs with Graph
  Reasoning Ability via Prompt Augmented by ChatGPT}.
\newblock \bibinfo{journal}{\emph{arXiv preprint arXiv:2304.11116}}
  (\bibinfo{year}{2023}).
\newblock


\bibitem[Zhao et~al\mbox{.}(2023)]%
        {39}
\bibfield{author}{\bibinfo{person}{Wayne~Xin Zhao}, \bibinfo{person}{Kun Zhou},
  \bibinfo{person}{Junyi Li}, \bibinfo{person}{Tianyi Tang},
  \bibinfo{person}{Xiaolei Wang}, \bibinfo{person}{Yupeng Hou},
  \bibinfo{person}{Yingqian Min}, \bibinfo{person}{Beichen Zhang},
  \bibinfo{person}{Junjie Zhang}, \bibinfo{person}{Zican Dong},
  {et~al\mbox{.}}} \bibinfo{year}{2023}\natexlab{}.
\newblock \showarticletitle{A survey of large language models}.
\newblock \bibinfo{journal}{\emph{arXiv preprint arXiv:2303.18223}}
  (\bibinfo{year}{2023}).
\newblock


\bibitem[Zheng et~al\mbox{.}(2022)]%
        {18}
\bibfield{author}{\bibinfo{person}{Yingying Zheng}, \bibinfo{person}{Wensheng
  Dou}, \bibinfo{person}{Yicheng Wang}, \bibinfo{person}{Zheng Qin},
  \bibinfo{person}{Lei Tang}, \bibinfo{person}{Yu Gao}, \bibinfo{person}{Dong
  Wang}, \bibinfo{person}{Wei Wang}, {and} \bibinfo{person}{Jun Wei}.}
  \bibinfo{year}{2022}\natexlab{}.
\newblock \showarticletitle{Finding bugs in Gremlin-based graph database
  systems via randomized differential testing}. In
  \bibinfo{booktitle}{\emph{Proceedings of the 31st ACM SIGSOFT International
  Symposium on Software Testing and Analysis}}. \bibinfo{pages}{302--313}.
\newblock


\end{thebibliography}

	%
	%
	%
	%
	%
	%
	%
	%
	
\end{document}